\documentstyle[epsfig,sprocl]{article}

\def \bea{\begin{eqnarray}}
\def \beq{\begin{equation}}
\def \bo{B^0}

\def \dz{D^0}
\def \eea{\end{eqnarray}}
\def \eeq{\end{equation}}

\def \hp{\hat{p}}
\def \ket#1{| #1 \rangle}
\def \ko{K^0}
\def \mat#1#2{\langle #1 | #2 \rangle}
\def \ob{\overline{B}^0}
\def \od{\overline{D}^0}
\def \of{\overline{f}}
\def \ok{\overline{K}^0}
\def \ot{\overline{t}}
\def \pr{\parallel}
\def \s{\sqrt{2}}
\def \st{\sqrt{3}}
\def \sx{\sqrt{6}}
\def \tl{\tilde{\lambda}}
\begin{document}

\title{CP VIOLATION IN $B$ DECAYS
}
% $^*$}

\author{JONATHAN L. ROSNER}

\address{Enrico Fermi Institute and Department of Physics, University of
Chicago \\ 5640 South Ellis Avenue, Chicago, IL 60637 \\ 
E-mail: rosner@hep.uchicago.edu} 

\maketitle

\abstracts{The role of $B$ decays in the study of CP violation is reviewed.
We treat the interactions and spectroscopy of the $b$ quark and then
introduce CP violation in $B$ meson decays, including time-dependences, decays
to CP eigenstates and non-eigenstates, and flavor tagging.  Additional topics
include studies of strange $B$'s, decays to pairs of light pseudoscalar
mesons, and the roles of gluonic and electroweak penguin diagrams, and
final-state interactions.}

\section{Introduction}

Discrete symmetries such as time reversal (T), charge conjugation
(C), and space inversion or parity (P) have provided both clues and puzzles
in our understanding of the fundamental interactions.  The realization that
the charge-changing weak interactions violated P and C maximally was central to
their formulation in the $V-A$ theory.  The
theory was constructed in 1957 to conserve the product CP, but within seven
years the discovery of decay of the long-lived neutral kaon to two pions
\cite{CCFT} showed that even CP was not conserved.  Nearly twenty years later,
Kobayashi and Maskawa (KM) \cite{KM} proposed that CP violation in the neutral
kaon system could be explained in a model with three families of quarks, at a
time (1973) when no evidence for the third family and not even all evidence
for the second had been found.  The quarks of the third family, now denoted by
$b$ for bottom and $t$ for top, were subsequently discovered in 1977
\cite{ups} and 1994, \cite{top} respectively.

Decays of hadrons containing $b$ quarks now appear to be particularly
fruitful ground for testing the KM hypothesis and for
displaying evidence for any new physics beyond this ``standard model'' of
CP violation.  A meson containing a $\bar b$ quark will be known generically
as a $B$ meson, in the same way as a $K$ meson contains an (anti-) strange
quark $\bar s$.  The present lectures are devoted to some tests of CP
violation utilizing $B$ meson decays.  (Baryons containing $b$ quarks also
may display CP violation but we will not discuss them here.)

We first deal with the spectroscopy and interactions
of the $b$ quark.  In Section 2 we describe the discovery of the charmed quark,
the tau lepton, the $b$ quark, and $B$ mesons.  Section 3 is devoted to the
spectroscopy of hadrons containing the $b$ quark, while Section 4 treats its
weak interactions.  Neutral mesons containing the $b$ quark can mix with
their antiparticles (Section 5), providing important information on the weak
interactions of $b$ quarks.

We then introduce CP violation in $B$ meson decays.
After general remarks and a discussion of decays to CP eigenstates
(Section 6) we turn to decays to CP-noneigenstates (Section 7) and describe
various methods of tagging the flavor of an initially-produced $B$ meson
(Section 8).  Some specialized topics
include strange $B$'s (Section 9), decays to pairs of light mesons (Section
10), and the roles of penguin diagrams (Section 11), and final-state
interactions (Section 12).

Topics not covered in detail in the lectures but worthy of mention in this
review are noted briefly in Section 13.  The possibility that the Standard
Model of CP violation might fail at some future time to describe all the
observed phenomena is discussed in Section 14, while Section 15 concludes.
Other contemporary reviews of the subject \cite{YN,RF} may be consulted.

\section{Discovery of the $b$ quark}

\subsection{Prelude: The charmed quark}

During the 1960's and 1970's, when the electromagnetic and weak interactions
were being unified by Glashow, Weinberg, and Salam,\cite{GWS} it was
realized \cite{CQ} that a consistent theory of hadrons required a parallel
\cite{qu} between the then-known two pairs of weak isodoublets of leptons,
$(\nu_e,e^-),~(\nu_\mu,\mu^-)$,
and a corresponding multiplet structure for quarks,
$(u,d),~(c,s)$.
The known quarks at that time consisted of one with charge 2/3, the up
quark $u$, and two with charge $-1/3$, the down quark $d$ and the
strange quark $s$.  The charmed quark $c$ was a second quark with charge 2/3
and a proposed mass of about 1.5 to 2 GeV/$c^2$.\cite{GIM,GLR}

The parallel between leptons and quarks
was further motivated by the cancellation of anomalies \cite{CQ,BIM}
in the electroweak theory.  These are associated with triangle graphs
involving fermion loops and three electroweak currents.
It is sufficient to consider the anomaly for the product $I_{3L} Q^2$, where
$I_{3L}$ is the third component of left-handed isospin and $Q$ is the
electric charge.  The sum $\sum_i (I_{3L})_i Q_i^2$ over all fermions $i$
must vanish.  If a family of quarks and leptons consists of one weak
isodoublet of quarks and one of leptons, this cancellation can be implemented
within a family, as illustrated in Table \ref{tab:anom}.
 
\begin{table}[t]
\caption{Anomaly cancellation in the electroweak theory. \label{tab:anom}}
\vspace{0.2cm}
\begin{center}
\footnotesize
\begin{tabular}{|l|c|c|c|c|} \hline
Family         &   1   &    2    &    3     & Contribution per family \\ \hline
Neutrino  & $\nu_e$ & $\nu_\mu$ & $\nu_\tau$ & $(1/2)(0)^2 = 0$ \\
Charged lepton & $e^-$ & $\mu^-$ & $\tau^-$ & $(-1/2)(-1)^2 = -1/2$ \\
$Q=2/3$ quark  & $u$   &   $c$   &   $t$    & $3(1/2)(2/3)^2 = 2/3$ \\
$Q=-1/3$ quark & $d$   &   $s$   &   $b$    & $3(-1/2)(-1/3)^2 = -1/6$ \\
\hline
\end{tabular}
\end{center}
\end{table}

The first hints of charm arose in nuclear emulsions \cite{Niu} and were
recognized as such by Kobayashi and Maskawa.\cite{KM}  However, more
definitive evidence appeared in November, 1974, in the form of the $^3S_1$
$c \bar c$ ground state discovered simultaneously on the East \cite{J}
and West \cite{psi} Coasts of the U.\ S.\ and named, respectively,
$J$ and $\psi$.  The East Coast experiment utilized the reaction $p +
{\rm Be} \to e^+ e^- + \ldots$ and observed the $J$ as a peak at 3.1 GeV$/c^2$
in the effective $e^+ e^-$ mass.  The West Coast experiment studied $e^+
e^-$ collisions in the SPEAR storage ring and saw a peak in the cross section
for production of $e^+ e^-$, $\mu^+ \mu^-$, and hadrons at a center-of-mass
energy of 3.1 GeV.  Since the discovery of the $J/\psi$ the charmonium level
structure has blossomed into a richer set of levels than has been observed
for the original ``onium'' system, the $e^+ e^-$ positronium bound states.

The lowest charmonium levels are narrow because they are kinematically
unable to decay to pairs of charmed mesons (each containing a single charmed
quark).  The threshold for this decay is at a mass of about 3.73 GeV$/c^2$.
Above this mass, the charmonium levels gradually become broader.  The charmed
mesons, discovered in 1976 and subsequently, include $D^+ = c \bar d$
(mass 1.869 GeV/$c^2$), $D^0 = c \bar u$ (mass 1.865 GeV/$c^2$), and $D_s = c
\bar s$ (mass 1.969 GeV/$c^2$).  These mesons were initially hard to find
because the large variety of their possible decays made any one mode
elusive.  For example, the two-body decay $D^0 \to K^- \pi^+$ has a branching
ratio of only about 3.8\%; \cite{PDG} higher-multiplicity decays are somewhat
favored.

\subsection{Prelude:  The $\tau$ lepton}

About the same time as the discovery of charm, another signal was showing
up in $e^+ e^-$ collisions at SPEAR, corresponding to the production of
a pair of new leptons:  $e^+ e^- \to \gamma^* \to \tau^+ \tau^-$.\cite{tau}
The $\tau$ signal had a number of features opposite to those of charm:
lower- rather than higher-multiplicity decays and fewer rather than more
kaons in its decay products, for example, so separating the two contributions
took some time.\cite{HH}

The mass of the $\tau$ is 1.777 GeV/$c^2$.  Its favored decay products are
a tau neutrino, $\nu_\tau$, and whatever the charged weak current can
produce, including $e \bar \nu_e$, $\mu \bar \nu_\mu$, $\pi$, $\rho$, etc.
It thus contributes somewhat less than one unit to
\beq
R \equiv \sum_i Q_i^2 = \frac{\sigma(e^+ e^- \to {\rm hadrons})}
{\sigma(e^+ e^- \to \mu^+ \mu^-)}~~~,
\eeq
which would have risen from the value of 2 for $u,d,s$ quarks below
charm threshold to 10/3 above charm threshold if charm alone were being
produced, but was seen to rise considerably higher.

One problem with accepting the $\tau$ as a companion of the charmed quark
was that the neat anomaly cancellation provided by the charmed quark,
mentioned above, was immediately upset.  The anomaly contributed by the
$\tau$ lepton would have to be cancelled by further particles, such as a pair
of new quarks $(t,b)$ with charge 2/3 and $-1/3$.  Such quarks had indeed
already been utilized two years before the $\tau$ was established, in 1973
by Kobayashi and Maskawa \cite{KM} in their theory of CP violation.  The
names ``top'' and ``bottom'' were coined by Harari in 1975,\cite{HHS} in
analogy with ``up'' and ``down.''
 
\subsection{Dilepton spectroscopy}

One reason for the experiment which discovered the $J$ particle \cite{J}
was an earlier study, also at Brookhaven National Laboratory, by L. Lederman
and his collaborators, of $\mu^+ \mu^-$
pairs produced in proton-uranium collisions.\cite{LLJ}  The $m(\mu^+ \mu^-)$
spectrum in this experiment displayed a shoulder around 3.5 GeV$/c^2$.  It
was not recognized as a resonant peak and was displaced in mass from the true
$J/\psi$ value because of the poor mass resolution of the experiment.

After the discovery of the $J/\psi$, Lederman's group continued to pursue
dilepton spectroscopy.  In 1977 a search with greater
sensitivity and better mass resolution turned up evidence for peaks at
9.4, 10.0, and possibly 10.35 GeV$/c^2$.\cite{ups}  These were candidates
for the 1S, 2S, and 3S $^3S_1$ levels of a new $Q \bar Q$ system.  Several
pieces of evidence identified the heavy quark $Q$ as a $b$ quark.

(1) The $\Upsilon$(1S) and $\Upsilon'$(2S) were produced in 1978 by the
electron-positron collider DORIS at DESY and their partial widths to
$e^+ e^-$ pairs were measured.\cite{DORIS}  It was shown \cite{QRQ} that if the
$Q \bar Q$ system was bound by the same quantum
chromodynamic force as as the $c \bar c$ (charmonium) system, one could use
the $c \bar c$ states to gain some idea about the details of the $Q \bar Q$
binding.  Since $\Gamma(Q \bar Q) \propto e_Q^2$, where $e_Q$ is the charge
of the quark $Q$, it was possible to conclude from the data that $|e_Q|=1/3$
was favored over $|e_Q| = 2/3$.

(2) The Cornell $e^+ e^-$ ring CESR began operating in 1979,\cite{CESR}
reaching a fourth $\Upsilon$(4S) peak and finding it broader than the
first three.  This indicated that the meson pair threshold lay below
$M[\Upsilon$(4S)]$ = 10.58$ GeV/$c^2$.  Farther above this threshold, wiggles
in the total cross section for hadron production averaged out to indicate
a step in $R$ of 1/3, confirming that $|e_Q| = 1/3$.

(3) The possibility that $Q$ was an isosinglet quark of charge $-1/3$, and
thus not the partner of some quark $t$ with charge 2/3, was
ruled out by the absence of significant flavor-changing neutral current
decays such as $b \to s \mu^+ \mu^-$.\cite{KP,CESRNC}

The structure of the $\Upsilon$ levels
is remarkably similar to that of the charmonium levels except for having
more levels below flavor threshold.  For example, the fact that the 3S level
is below flavor threshold allows it to decay to the 2P levels via electric
dipole transitions with appreciable branching ratios; the transitions between
the S and P levels are well described in potential models which reproduce other
aspects of the spectra.  Several reviews treat the fascinating
regularities of the spectroscopy of these levels.\cite{specrevs}

\subsection{Discovery of $B$ mesons}

The lightest meson containing a $b$ quark and each flavor of light antiquark
is expected to decay weakly.  The allowed decays of $b$ are $(c~{\rm or}~u)~+$
(virtual $W^-$), with the $c$ giving rise to lots of strange particles while
the $u$ gives few strange particles.  The virtual $W^-$ can decay to $\bar u
d$, $\bar c s$, $e^- \bar \nu_e$, $\mu^- \bar \nu_\mu$, and $\tau^- \bar
\nu_\tau$.

In $e^+ e^-$ collisions above $B \bar B$ threshold, several signals of $B$
meson production were observed by the CLEO Collaboration starting around
1980: \cite{ET}

\begin{itemize}

\item Prompt leptons (signals of semileptonic decay)

\item An abundance of kaons (a signal that $b \to c + W^-_{\rm virt}$ is
preferred over $b \to u + W^-_{\rm virt}$)

\item ``Daughter'' (lower-momentum) leptons from $c$ semileptonic decays.

\end{itemize}

These indirect signals were followed by reconstruction of
$B^+$ and $\bo$ decays,\cite{Brecon} e.g.,
\beq
B^+ = \bar b u \to \bar c u \bar d u \to \bar D^0 \pi^+~~,~~~
\bo = \bar b d \to \bar c u \bar d d = D^- \pi^+~~~.
\eeq
Typical branching ratios for these final states \cite{PDG} are $(5.3 \pm 0.5)
\times 10^{-3}$ for $\bar D^0 \pi^+$ and $(3.0 \pm 0.4) \times 10^{-3}$ for
$D^- \pi^+$. $\bo \to \bar D^0 \pi^0$ is also allowed but
not yet observed.  These small branching
ratios mean that reconstruction of exclusive final states is even harder
for $B$ mesons than for charmed particles.

\section{The known $B$ hadrons}

\subsection{$B$ mesons}

\begin{table}
\caption{Ground-state heavy-light $(Q \bar q)$ pseudoscalar mesons and the
corresponding vector mesons.
Here the spectroscopic notation $^{2L+1}L_J$ is used to denote the spin,
orbital, and total angular momenta of the $Q \bar q$ state.
\label{tab:Bmes}}
\begin{center}
\begin{tabular}{|c|c|c|c|c|} \hline
 & \multicolumn{2}{c}{Pseudoscalar ($^1S_0$) meson|} &
   \multicolumn{2}{c}{Vector ($^3S_1$) meson|} \\ \hline
Quark content & Name & Mass (GeV/$c^2$) & Name & Mass (GeV/$c^2$) \\ \hline
$c \bar u$ & $D^0$ & $ 1864.5 \pm 0.5$ & $D^{*0}$ & $ 2006.7 \pm 0.5$ \\
$c \bar d$ & $D^+$ & $ 1869.3 \pm 0.5$ & $B^{*+}$ & $ 2010.0 \pm 0.5$ \\
$c \bar s$ & $D_s$ & $ 1968.6 \pm 0.6$ & $D_s^*$  & $ 2112.4 \pm 0.7$ \\ \hline
$\bar b u$ & $B^+$ & $5279.0 \pm 0.5$ & $B^{*+}$ & $5325.0 \pm 0.6$ \\
$\bar b d$ & $B^0$ & $5279.4 \pm 0.5$ & $B^{*0}$ & $5325.0 \pm 0.6$ \\
$\bar b s$ & $B_s$ & $5369.6 \pm 2.4$ & $B_s^*$  & $\simeq 5416$ \\ \hline
\end{tabular}
\end{center}
\end{table}

The nonstrange ground-state $B$ (pseudoscalar) and $B^*$ (vector) mesons
are compared with the corresponding charmed mesons in Table \ref{tab:Bmes}.
Evidence for the $B_s^*$ exists in the form of a photon signal for the
decay $B_s^* \to B_s \gamma$.\cite{CUSBBs}  The photon energy, 46 MeV, is
expected to be the same as that seen in $B^{*0} \to B^0 \gamma$.\cite{RW}

Since the $B^*$ and $B$ states are separated by only 46 MeV, a $B^*$
should always decay to a $B$ of the same flavor and a photon.
This is in contrast to the case of the $D^*$ and $D$ states, whose separation
is just about a pion mass.  The electromagnetic mass splittings are such that
$D^{*+} \to D^0 \pi^+$, $D^{*+} \to D^+ \pi^0$, and $D^{*0} \to D^0 \pi^0$
are just barely allowed, while $D^{*0} \to D^+ \pi^-$ is forbidden.  The
low-momentum $\pi^+$ in $D^{*+} \to D^0 \pi^+$ acts as a ``tag,'' useful
both for signalling the production of a charmed meson \cite{Nus} and, by its
charge, distinguishing the $D^0$ from a $\bar D^0$.  Since $B^*$ decays are
not useful for this type of ``flavor'' tag, one must resort to the decays of
heavier excited $b \bar q$ states (Section 8).

The hyperfine splitting of $B$ mesons is smaller than that in charmed mesons
because the chromomagnetic moments of the heavy quarks scale as the
inverse of their masses:
\beq
\frac{m_{B^*} - m_B}{m_{D^*} - m_D} \simeq \frac{1/m_b}{1/m_c} =
\frac{m_c}{m_b} \simeq \frac{1}{3}~~~.
\eeq

\subsection{The $\Lambda_b$ baryon}

The lightest baryon containing a $b$ quark is the $\Lambda_b = b[ud]_{I=0}$.
Its mass is $5624 \pm 9$ MeV$/c^2$.\cite{PDG}  The $ud$ system must be in a
color $3^*$ (antisymmetric) state, since the $b$ is a color triplet and the
$\Lambda_b$ is a color singlet.  The spin-zero state of $ud$ is favored over
the spin-one state by the chromodynamic hyperfine interaction.  By Fermi
statistics, the $ud$ pair must then be in an (antisymmetric) isospin-zero 
state.  For similar reasons, the $I=0$ state of a strange quark and two
nonstrange quarks, the $\Lambda = s[ud]_{I=0}$ with mass 1116 MeV$/c^2$, is
lighter than the $\Sigma = s(uu,ud,dd)_{I=1}$ states with average mass
1193 MeV$/c^2$.

The charmed analogue of the $\Lambda_b$ is the $\Lambda_c = c[ud]_{I=0}$ with
mass $2284.9 \pm 0.6$ MeV/$c^2$. The difference in mass of the two particles is
$M(\Lambda_b) - M(\Lambda_c) = 3339 \pm 9$ MeV$/c^2$.  This provides 
an estimate of $m_b - m_c$ since there are no hyperfine terms involving
the heavy quark; the light-quark system has zero spin in both baryons.  There
will be a correction of order $m_c^{-1} - m_b^{-1}$ due to possible differences
in kinetic energies.

One can perform a similar estimate for $Q \bar q$ mesons by eliminating the
hyperfine energy, performing a suitable average over vector ($^3S_1$) and
pseudoscalar ($^1S_0$) meson masses.  The expectation value of the relevant
interaction
term is $\langle \sigma_Q \cdot \sigma_{\bar q} \rangle = (1,-3)$ for
$(^3S_1,^1S_0)$ states.  Thus the hyperfine energy is absent in the combination
$[3M(^3S_1) + M(^1S_0)]/4$.  Since $(3M_{D^*} + M_D)/4
= 1973$ MeV/$c^2$ and $(3M_{B^*} + M_B)/4 = 5313$ MeV/$c^2$ (taking isospin
averages), we estimate from the mesons that $m_b - m_c = 3340$ MeV$/c^2$,
identical to the estimate from the baryons.  (Again, kinetic energies could
provide a correction to this result.)  The spectroscopic assignment of the
$\Lambda_b$ is thus likely to be the correct one.

Known decay modes of the $\Lambda_b$ include
$\Lambda_c^+ \ell^- \bar \nu_\ell$ and
$J/\psi \Lambda$.  The fact that the corresponding quark subprocesses
$b \to c \ell^- \bar \nu_\ell$ and $b \to c \bar c s$ conserve isospin leads
to the requirement that the final states have zero isospin in both cases,
restricting the number of additional pions that can be produced.

\section{Interactions of the $b$ quark}

In this Section we shall discuss the way in which the interactions of the
$b$ quark provide information on the pattern of charge-changing weak
interactions of quarks parametrized by the Cabibbo-Kobayashi-Maskawa (CKM)
matrix $V$.\cite{KM,Cab}  More details on determination of the CKM matrix are
included in the lectures by Buchalla,\cite{Buch} DeGrand,\cite{DeG}
Falk,\cite{Falk}, Neubert,\cite{NeuTASI} and Wolfenstein,\cite{Wolf} 

\subsection{The $b$ lifetime:  indication of small $|V_{cb}|$}

The long $b$ quark lifetime ($> 1$ ps) indicated that the CKM element $V_{cb}$
was considerably smaller than $|V_{us}| \simeq |V_{cd}| \simeq 0.22$.  One
can estimate $V_{cb}$ using a free-quark method.

The subprocess $b \to c W^{*-} \to c \ell^- \bar \nu_\ell$ has a rate
\beq
\Gamma(b \to c \ell^- \bar \nu_\ell) = \frac{G_F^2}{192 \pi^3} m_b^5
|V_{cb}|^2 f(m_b,m_c,m_\ell)~~~,
\eeq
where $G_F = 1.16637(2) \times 10^{-5}$ GeV$^{-2}$ is the Fermi coupling
constant.  In the limit in which
the lepton mass can be neglected, $f(m_b,m_c,m_\ell) = f(m_c^2/m_b^2)$, with
$f(x) = 1 - 8x + 8x^3 - x^4 - 12 x^2 \ln x$.  The uncertainty in the prediction
for $\Gamma(b \to c \ell^- \bar \nu_\ell)$ due to that in $m_b$ is
mitigated by the constraint noted above on $m_b - m_c \simeq 3.34$ GeV/$c^2$.

Taking a nominal range of quark masses around $m_b = 4.7$ GeV/$c^2$
(and hence a range around $m_c = 1.36$ GeV/$c^2$, $f(m_c^2/m_b^2)
= 0.54$), $\tau_b = 1.6 \times 10^{-6}$ s,\cite{PDG} and the branching ratio
${\cal B}(b \to c \ell \nu_\ell) \simeq 10.2\%$, one finds
\beq
|V_{cb} \simeq 0.0384 - 0.0008 \left( \frac{m_b - 4.7~{\rm GeV}/c^2}{0.1
~{\rm GeV}/c^2} \right)~~~.
\eeq
Thus if $m_b$ is uncertain by 0.3 GeV/$c^2$ (my guess), $|V_{cb}|$ is
uncertain by $\pm 0.0024$.  Recent averages \cite{PDG,Falk} give rise to
values of $|V_{cb}|$ somewhat above 0.040 with errors of $\pm 0.002$ to
$\pm 0.003$.

A new report by the CLEO Collaboration \cite{CLEOVcb} finds
$|V_{cb}| = 0.0462 \pm 0.0036$ based on the exclusive decay
process $B^0 \to D^{*-} \ell^+ \nu_\ell$.  This new determination bears
watching as it would affect many conclusions regarding
predictions for CP-violating asymmetries in $B$ decays.
We shall take $|V_{cb}| = 0.041 \pm 0.003$ as representing a
conservative range of present values.

\subsection{Charmless $b$ decays:  indication of smaller $|V_{ub}|$}

Although the $u$ quark is lighter than the $c$ quark, its production in
$b$ decays is disfavored, with $\Gamma(b \to u \ell \nu)/\Gamma(b \to c
\ell \nu)$ only about 2\%.  Since the phase-space factor $f(m_u^2/m_b^2)$
is very close to 1, while $f(m_c^2/m_b^2) \simeq 1/2$, this means that
$|V_{ub}/V_{cb}|^2 \simeq 1\%$, or $|V_{ub}/V_{cb}| \simeq 0.1$.  The
error on this quantity is dominated by theoretical uncertainty \cite{Falk};
detailed studies \cite{Flg} indicate $|V_{ub}/V_{cb}| = 0.090 \pm 0.025$.

\subsection{Pattern of charge-changing weak quark transitions}

\begin{figure}
\centerline{\epsfysize=3in \epsffile{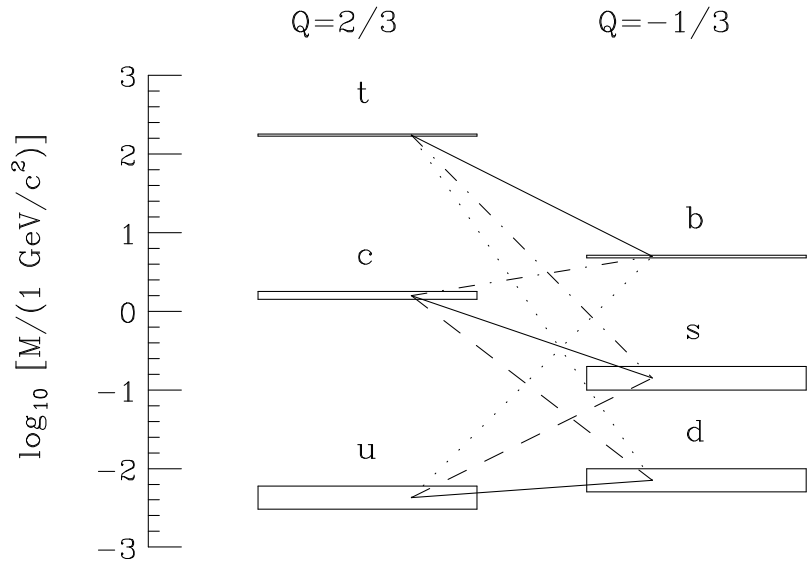}}
\caption{Pattern of charge-changing weak transitions among quarks.  Solid
lines:  relative strength 1; dashed lines:  relative strength 0.22; 
dot-dashed lines:  relative strength 0.04; dotted lines:  relative strength
$\le 0.01$. Breadths of lines denote estimated errors. \label{fig:trans}}
\end{figure}

The relative strengths of charge-changing weak quark transitions are
illustrated in Fig.\ \ref{fig:trans}.  {\it Why} the pattern looks like this is
a mystery, one of the questions (along with the values of the quark masses)
to be answered at a deeper level.

The interactions in Fig.\ \ref{fig:trans} may be parametrized by a
Cabibbo-Kobayashi-Maskawa (CKM) matrix of the form \cite{WP}
\beq
V_{\rm CKM} = \left[ \begin{array}{c c c}
1 - \frac{\lambda^2}{2} & \lambda & A \lambda^3 (\rho - i \eta) \\
- \lambda & 1 - \frac{\lambda^2}{2} & A \lambda^2 \\
A \lambda^3 (1 - \rho - i \eta) & - A \lambda^2 & 1 \end{array}
\right]~~~.
\eeq
The columns refer to $d,s,b$ and the rows to $u,c,t$.  The parameter
$\lambda = 0.22$ represents $\sin \theta_c$, where $\theta_c$ is the
Gell-Mann--L\'evy--Cabibbo \cite{Cab,GL} angle.  The value $|V_{cb}| =
0.041 \pm 0.003$ indicates $A = 0.85 \pm 0.06$, while $|V_{ub}/V_{cb}| = 0.090
\pm 0.025$ implies $(\rho^2 + \eta^2)^{1/2} = 0.41 \pm 0.11$.

Further information may be obtained by assuming that box diagrams involving
internal quarks $u,c,t$ with charge 2/3 are responsible for both the
CP-violating contribution to $K^0$--$\bar K^0$ mixing and to mixing between
neutral $B$ mesons and their antiparticles.  The parameter $|\epsilon_K|
= (2.27 \pm 0.02) \times 10^{-3}$ (see Buchalla's lectures \cite{Buch}) then
implies a constraint \cite{JRCKM}
\beq
\eta(1 - \rho + 0.39) = 0.35 \pm 0.12~~~,
\eeq
where the $1 - \rho$ term in parentheses arises from box diagrams with two
internal top quarks, while the correction 0.39 is due to diagrams with one
charmed and one top quark.  The error on the right-hand side is due primarily
to uncertainty in the Wolfenstein parameter $A = |V_{cb}|/\lambda^2$, which
enters to the fourth power in the $t \bar t$ contribution to $\epsilon_K$.
A lesser source of error is uncertainty in the parameter $B_K$ describing the
quark box diagram's matrix element between a $K^0$ and a $\bar K^0$.  We have
chosen \cite{Lubicz} $B_K = 0.87 \pm 0.13$.

% This is Figure 1

\begin{figure}
\centerline{\epsfysize=2.5in \epsffile{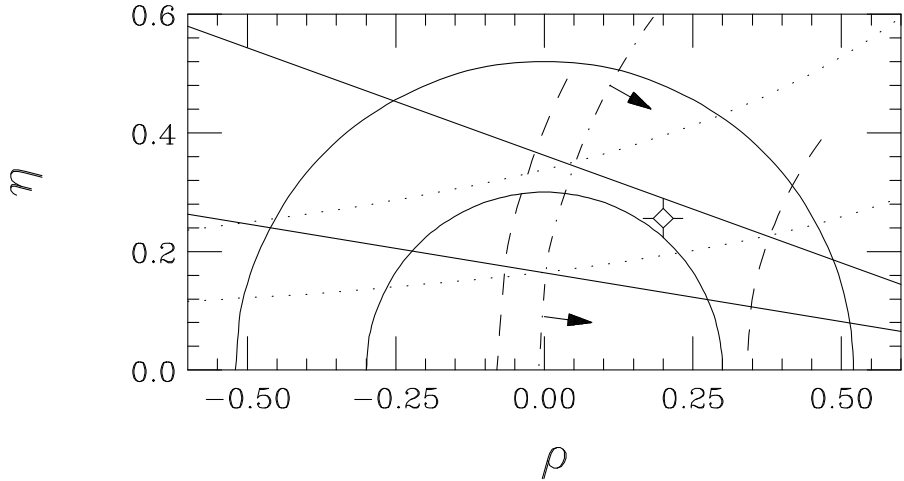}}
\caption{Region of $(\rho,\eta)$ specified by $\pm 1 \sigma$ constraints on
CKM matrix parameters. Solid semicircles denote limits
based on $|V_{ub}/V_{cb}| = 0.090 \pm 0.025$; dashed arcs denote limits
$|1 - \rho - i \eta|$ based on $\bo$--$\ob$ mixing; dot-dashed arc
denotes limit $|1 - \rho - i \eta| < 1.01$ based on $B_s$--$\overline{B_s}$
mixing; dotted lines denote limits $\eta (1 - \rho + 0.39) = 0.35 \pm 0.12$
based on CP-violating $\ko$--$\ok$ mixing.  Rays:  $\pm 1 \sigma$ limits
on $\sin 2 \beta$ (see Sec.\ 6.4).  The plotted point at $(\rho,\eta)
%U              |
\simeq (0.20,0.26)$ lies roughly in the middle of the allowed
region.  \label{fig:re}}
\end{figure}

Present information on $\bo$--$\ob$ mixing, interpreted in terms of
box diagrams with two quarks of charge 2/3, leads to a constraint on
$|V_{td}|^2$ which implies \cite{JRCKM} $|1- \rho - i \eta| =0.87 \pm 0.21$ for
the parameter range $f_B \sqrt{B_B} = 230 \pm 40$ MeV describing the matrix
element of the short-distance 4-quark operator taking $\bar b d$ into $\bar d
b$ between $\bo$ and $\ob$ states.  The best lower limit on
$B_s^0$--$\overline{B_s}^0$ mixing,\cite{Bslim} $\Delta M_s > 15$ ps$^{-1}$,
when compared with the corresponding value for $\bo$--$\ob$ mixing,
$\Delta m_d = 0.487 \pm 0.014$ ps$^{-1}$, leads to the bound
\beq
\frac{f_{B_s}^2 B_{B_s}}{f_B^2 B_B} \left| \frac{V_{ts}}{V_{td}} \right|^2 > 29
~~~.
\eeq
This may be combined with the estimate \cite{JRFM} $f_{B_s} \sqrt{B_{B_s}} \le
1.25 f_B \sqrt{B_B}$ based on quark models.  (Lattice gauge theories \cite{DeG} 
estimate this coefficent more precisely, generally giving values between
1.1 and 1.2.)  One finds $|V_{ts}/V_{td}| \ge 4.4$ or $|1 - \rho - i \eta| <
1.01$.  The constraints may be combined to yield the
allowed range in $(\rho,\eta)$ space illustrated in Fig.\ \ref{fig:re}.
Smaller regions are quoted in other reviews \cite{CKMrevs} which
view the theoretical sources of error differently.

% This is Figure 2

\subsection{Unitarity triangle}

The unitarity of the CKM matrix implies that to the order we are considering,
$V^*_{ub} + V_{td} = A \lambda^3$.  If this equation is divided by
$A \lambda^3$, one obtains a triangle in the $(\rho,\eta)$ complex plane
whose vertices are at $(0,0)$ (internal angle $\gamma$), $(\rho,\eta)$
(internal angle $\alpha$), and $(1,0)$ (internal angle $\beta$).

CP-violating asymmetries in certain $B$ decays can measure such quantities as
$\sin 2 \alpha$ and $\sin 2 \beta$.  The former, measured in $\bo \to \pi^+
\pi^-$ with some corrections due to ``penguin'' diagrams, may occupy a wide
range, as illustrated in Fig.\ \ref{fig:re}.  The latter, measured in the
``golden mode'' $\bo \to J/\psi K_S$ with few uncertainties, is more
constrained by other observables.

The goal of measurements of CP violation and other quantities in $B$ decays
will be to test the consistency of this picture and to either restrict the
parameter space further, thus providing a reliable target for future theories
of these parameters, or to expose inconsistencies that will point to new
physics.  Hence part of the program will be to overconstrain the unitarity
triangle, measuring both sides and angles in several different types of
processes.  While we discuss such measurements based on $B$ mesons,
Buchalla \cite{Buch} describes how, for example, $K^+ \to \pi^+ \nu \bar
\nu$ constrains the combination $|1 - \rho - i \eta + 0.44|$, where the
last term is a charmed quark correction to the dominant top quark contribution,
and the purely CP-violating process $K_L \to \pi^0 \nu \bar \nu$ constrains
$\eta$.

\section{Mixing of neutral $B$ mesons}

\subsection{Mass matrix formalism}

We shall work in a two-component basis utilizing the states $(\bo, \ob)$.
[It is also sometimes useful to consider a basis \cite{Wolf} $(B_+,B_-)$, where
$B_\pm = (\bo \pm \ob)/\s$.]  The time-dependence of these states is
described via a {\it mass matrix} ${\cal M} = M - i \Gamma/2$, where $M$ and
$\Gamma$ are Hermitian by definition:
\beq
i \frac{\partial}{\partial t} \left[ \begin{array}{c} \bo \\ \ob \end{array}
\right] = {\cal M} \left[ \begin{array}{c} \bo \\ \ob \end{array} \right]~~~.
\eeq
The requirement of CPT invariance, which we shall assume henceforth,
implies ${\cal M}_{11} = {\cal M}_{22}$, or equal transition amplitudes for
$\ko \to \ko$ and $\ok \to \ok$.  {\it Exercise: (a) Show this. Remember time
reversal is an antiunitary operator.  (b) Show that a similar argument applied
to ${\cal M}_{12}$ or ${\cal M}_{21}$ leads to no constraint.  (c) Relate the
result in (a) to the result quoted by Wolfenstein \cite{Wolf} for the
$(\bo \pm \ob)/\s$ basis.}

[Answer to Part (a):  Insert the unit operator $(CPT)^{-1} CPT$ before and
after ${\cal M}$ in ${\cal M}_{11} = \mat{\bo}{{\cal M} | \bo}$.
Note that $CPT(M - i \Gamma/2)(CPT)^{-1} = M + i \Gamma/2$.  Then
$$
{\cal M}_{11} = \langle CPT \bo | M + \frac{i \Gamma}{2} | CPT \bo \rangle^*
$$
\beq
= \langle \ob | M + \frac{i \Gamma}{2} | \ob \rangle^*
= \langle \ob | M - \frac{i \Gamma}{2} | \ob \rangle = {\cal M}_{22}~~~,
\eeq
where the antiunitarity of T has been used in the first step.  For a
discussion of antiunitary operators see, e.g., Sakurai's book on
quantum mechanics.\cite{JJS}]

The eigenstates of ${\cal M}$ may be denoted by $B_H$ (``heavy'') and $B_L$
(``light''):
\beq
|B_{H,L} \rangle = p_{H,L} | \bo \rangle + q_{H,L} | \ob \rangle~~~.
\eeq
The corresponding eigenvalues $\mu_{H,L} = m_{H,L} - \frac{i}{2}\Gamma_{H,L}$
satisfy
\beq
{\cal M} \left[ \begin{array}{c} p_i \\ q_i \end{array} \right]
= \mu_i \left[ \begin{array}{c} p_i \\ q_i \end{array} \right]~~~(i=H,L)~~~,
\eeq
specifically,
\beq
{\cal M}_{11} + {\cal M}_{12} \frac{q_i}{p_i} = {\cal M}_{21} \frac{p_i}{q_i}
+ {\cal M}_{11} = \mu_i~~~,
\eeq
so that ${\cal M}_{12}(q_i/p_i) = {\cal M}_{21}(p_i/q_i)$, or $(p_i/q_i)^2
= {\cal M}_{12}/{\cal M}_{21}$.

For $\bo$--$\ob$ mixing, in contrast to the situation for neutral kaons,
the scarcity of intermediate states accessible to both $\bo$ and $\ob$ and
the presence of a large top-quark contribution to $M_{12}$ means that
$|\Gamma_{12}| \ll |M_{12}|$, so that
\beq \label{eqn:pq}
\frac{p_i}{q_i} = \pm \sqrt{\frac{M_{12}}{M_{21}}}~~~.
\eeq
We may choose $p_L = p_H = p$, $q_L = - q_H = q$.  Normalizing $|p|^2 + |q|^2
=1$, we then write
\beq
|B_L \rangle = p | \bo \rangle + q | \ob \rangle~~,~~~
|B_H \rangle = p | \bo \rangle - q | \ob \rangle~~~.
\eeq

The sign ambiguity in (\ref{eqn:pq}) may be resolved as follows.  Since
\beq
\mu_L = {\cal M}_{11} + {\cal M}_{12} \frac{q}{p} = {\cal M}_{11} +
 {\cal M}_{21} \frac{p}{q}~~~,
\eeq
\beq
\mu_H = {\cal M}_{11} - {\cal M}_{12} \frac{q}{p} = {\cal M}_{11} -
 {\cal M}_{21} \frac{p}{q}~~~,
\eeq
then $\mu_H - \mu_L = - 2 {\cal M}_{12}(q/p)$, which in the limit
$|\Gamma_{12}| \ll |M_{12}|$ is $\mu_H - \mu_L = (-,+) \sqrt{M_{12}M_{21}}$
for the choice of (+,--) in (\ref{eqn:pq}).  Since $M_{21} = M_{12}^*$
($M$ is Hermitian), we must take the -- sign in (\ref{eqn:pq}) in order that
the ``heavy'' mass $m_H$ be greater than the ``light'' mass $m_L$.  Then
\beq
\frac{p_i}{q_i} = - \sqrt{\frac{M_{12}}{M_{21}}}~~~.
\eeq

Neglecting $\Gamma_{12}$ in comparison with $|M_{12}|$, we then find
\beq
\Delta m \equiv m_{H} - m_{L} = 2 |M_{12}|~~,~~~
\Delta \Gamma \equiv \Gamma_H - \Gamma_L \simeq 0~~~.
\eeq
If one keeps $\Gamma_{12}$ to lowest order, one can show \cite{BaBarbk} that
\beq
\frac{q}{p} = - \frac{M_{12}^*}{|M_{12}|} \left[ 1 - \frac{1}{2} {\rm Im}
\left( \frac{\Gamma_{12}}{M_{12}} \right) \right]~~~.
\eeq
In the limit that $\Gamma_{12}$ is negligible and $\Delta \Gamma = 0$,
$q/p$ is a pure phase, determined by the phase of $M_{12}$.

Now, $M_{12}$ takes $\ob = b \bar d$ into $\bo = d \bar b$, so its phase
is that of $(V_{tb} V^*_{td})^2$, or $e^{2 i \beta}$.  Thus in this limit
we find $q/p \simeq e^{- 2 i \beta}$.  More specifically, in the phase
convention in which $(CP)|\bo \rangle = + | \ob \rangle$, we find
\beq
M_{12} = - \frac{G_F^2}{12 \pi^2} (V_{tb} V^*_{td})^2 M_W^2 m_B f_B^2 B_B
\eta_B S \left( \frac{m_t}{M_W} \right)~~~.
\eeq
Here $f_B$ is the $B$ meson decay constant, $B_B$ is the vacuum saturation
factor, $\eta_B = 0.55$ is a QCD correction factor,\cite{BJW} and \cite{IL}
\beq \label{eqn:S}
S(x) \equiv \frac{x}{4} \left[ 1 + \frac{3 - 9x}{(x-1)^2} + \frac{6 x^2 \ln
x}{(x-1)^3} \right]~~~.
\eeq
The appropriate top quark mass for this calculation \cite{BurasB} is
$m_t(m_t) \simeq 165$ GeV/$c^2$.  The BaBar Physics Book \cite{BaBarbk} may
be consulted for further conventions and details.

\subsection{Time dependences}

We would like to know how states which are initially $\bo$ or $\ob$ evolve in
time.  The mass eigenstates evolve as $B_i \to B_i e^{-i \mu_i t}~(i = L,H)$.
The flavor eigenstates are expressed in terms of them as
\beq
t=0:~~ \ket{\bo} = \frac{\ket{B_L} + \ket{B_H}}{2p}~~,~~~
       \ket{\ob} = \frac{\ket{B_L} - \ket{B_H}}{2q}~~~,
\eeq
\bea
t>0:~~
\ket{\bo(t)} & = & (\ket{B_L} e^{-i \mu_L t} + \ket{B_H} e^{-i \mu_H t})
/2p~~, \\
\ket{\ob(t)} & = & (\ket{B_L} e^{-i \mu_L t} - \ket{B_H} e^{-i \mu_H t})
/2q~~~.
\eea
Now substitute back for $B_{L,H}$:
\bea
\ket{\bo(t)} & = & \ket{\bo} f_+(t) + \frac{q}{p}f_-(t) \ket{\ob}~~~, \\
\ket{\ob(t)} & = & \ket{\ob} f_+(t) + \frac{p}{q}f_-(t) \ket{\bo}~~~,
\eea
\bea
f_+(t) & \equiv & e^{- i m t} e^{- \Gamma t/2} \cos(\Delta \mu t/2)~~~, \\
f_-(t) & \equiv & e^{- i m t} e^{- \Gamma t/2} i \sin(\Delta \mu t/2)~~~,
\eea
$\Delta \mu \equiv \mu_H - \mu_L = \Delta m - i (\Delta \Gamma/2)$,
$\Delta m \equiv m_H - m_L$, $\Delta \Gamma \equiv \Gamma_H - \Gamma_L$,
$m \equiv (m_H + m_L)/2$, $\Gamma \equiv(\Gamma_H + \Gamma_L)/2$.

Again, for simplicity, we shall neglect $\Delta \Gamma$ in comparison with
$\Delta m$.  A lowest-order quark model calculation (for which QCD
corrections change the answer) gives \cite{dg}
\beq \label{eqn:bgm}
\frac{\Gamma_{12}}{M_{12}} = -\frac{3 \pi}{2} \frac{m_t^2/M_W^2}{S(m_t^2/M_W^2)}
\frac{m_b^2}{m_t^2} \left( 1 + \frac{8}{3} \frac{m_c^2}{m_b^2}
\frac{V_{cb}V^*_{cd}}{V_{tb}V_{td}^*} \right) \simeq - \frac{1}{180}~~~,
\eeq
where $S(x)$ was defined in Eq.\ (\ref{eqn:S}). 
The intermediate states dominating the loop calculation of $\Gamma_{12}$
have typical mass scales $m_b$, whereas loop momenta of order $m_t$ give rise
to the main contributions to $M_{12}$.

Neglecting $\Delta \Gamma$ and performing time integrals, one finds
\beq
\Gamma \int_0^\infty \! dt |f_+(t)|^2 = \frac{2 + x_d^2}{2 (1 + x_d^2)}
~~,~~~ \Gamma \int_0^\infty \! dt |f_-(t)|^2 = \frac{x_d^2}
{2 (1 + x_d^2)}~~~,\eeq
where $x_d \equiv \Delta m_{B_d}/\Gamma_{B_d}$, and $B_d$ is another name
for $\bo = \bar b d$ to distinguish it from $B_s = \bar b s$.  The sum of the
two terms is 1.  The first term is 1 for $x_d = 0$, approaches 1/2 for $x_d
\to \infty$, and is about 0.82 for the actual value $x_d = 0.754 \pm 0.027$.
The second term is 0 for $x_d = 0$, approaches 1/2 for $x_d \to \infty$, and is
about 0.18 in actuality.  Thus a neutral non-strange $B$ of a given flavor
($\bo$ or $\ob$) has about 18\% probability of decaying as the opposite
flavor.

\section{CP violation}

\subsection{Asymmetry: general remarks}

We wish to compare $\mat{f}{\bo_{t=0}(t)}$ and $\mat{\of}{\ob_{t=0}(t)}$,
where $f$ is a final state and $\of \equiv (CP)f$.  Now define
\beq
x \equiv \frac{\mat{f}{\ob}}{\mat{f}{\bo}}~~,~~~
\bar x \equiv \frac{\mat{\of}{\bo}}{\mat{\of}{\ob}}~~,~~~
\lambda_0 \equiv \frac{q}{p}x~~,~~~
\bar \lambda_0 \equiv \frac{p}{q}\bar x~~~.
\eeq
Using the time evolution derived earlier for $\bo_{t=0}$ and $\ob_{t=0}$,
one then finds
\bea \label{eqn:td}
\mat{f}{\bo_{t=0}(t)} & = & \mat{f}{\bo} \left[ f_+(t) + \lambda_0(t) f_-(t)
\right]~~~,\\
\mat{\of}{\ob_{t=0}(t)} & = & \mat{\of}{\ob} \left[ f_+(t) + \bar
\lambda_0(t) f_-(t) \right]~~~.
\eea
This result can be simplified under several circumstances.  (a) If
there is a single strong eigenchannel, final-state strong interaction phases
in $x$ or $\bar x$ cancel, since the numerator and denominator refer to the
same final state.  Then $\bar x = x^*$, since weak phases flip sign under CP.
(b) Recall that $|q/p|$ is nearly 1 for $B$ mesons.  (For non-strange $B$'s,
we found $q/p \simeq e^{- 2 i \beta}$.)  Combining (a) and (b), we find
$\bar \lambda_0 = \lambda_0^*$ for these cases.

\subsection{Time-dependent asymmetry}

According to Eq.\ (\ref{eqn:td}), the rates for a $(\bo,\ob)$ produced at $t=0$
to evolve to the respective final states $(f,\of)$ at a time $t$ are
\bea
d  \Gamma(\bo_{t=0} \to f)/dt & \sim & |f_+(t) + \lambda_0 f_-(t)|^2~~~,\\
d  \Gamma(\ob_{t=0} \to \of)/dt & \sim & |f_+(t) + \bar \lambda_0
f_-(t)|^2~~~,
\eea
with the coefficients of proportionality identical if there is a single strong
eigenchannel.  Now consider the case of a CP-eigenstate $f$ such that $\of
= \pm f$.  Then we have not only $\bar x = x^*$ (see above), but also
$\bar x = x^{-1}$, so $|x| = 1$.  In that case, when $|q/p| = 1$ as is the
case for neutral $B$'s, we have $|\lambda_0| = 1$ and $\bar \lambda_0 =
\lambda_0^*$.  Then
\bea \label{eqn:rat}
|f_+ + \lambda_0 f_-|^2 & = & e^{-\Gamma t} \left|\cos \frac{\Delta m t}{2} +
i \lambda_0 \sin \frac{\Delta m t}{2} \right|^2 \\
& = & e^{-\Gamma t} \left[ 1 - {\rm Im} \lambda_0 \sin \Delta m t
\right]~~~,
\eea
\bea \label{eqn:tdm}
d \Gamma(\bo_{t=0} \to f)/dt & \sim & e^{-\Gamma t} \left[ 1 - {\rm Im}
\lambda_0 \sin \Delta m t \right]~~~, \nonumber \\
d  \Gamma(\ob_{t=0} \to \of)/dt & \sim &  e^{-\Gamma t} \left[ 1 + {\rm Im}
\lambda_0 \sin \Delta m t \right]~~~.
\eea
The second term in each of these equations consists of an exponential decay
modulated by a sinusoidal oscillation.  The time-dependent asymmetry is then
\beq
{\cal A}_f \equiv
\frac{d \Gamma(\bo_{t=0} \to f)/dt - d  \Gamma(\ob_{t=0} \to \of)/dt}
     {d \Gamma(\bo_{t=0} \to f)/dt + d  \Gamma(\ob_{t=0} \to \of)/dt}
= - {\rm Im} \lambda_0 \sin \Delta m t~~~.
\eeq
When $\Delta m/\Gamma \gg1$, the wiggles in Eqs.\ (\ref{eqn:tdm})
average out, and not much time-integrated asymmetry is possible,
while when $\Delta m/\Gamma \ll 1$, the decay occurs before there is time
for oscillations.  The maximal time-integrated asymmetry occurs when
$\Delta m/\Gamma = 1$.

When more than one eigenchannel is present, the condition $|\lambda_0| = 1$
need not be satisfied, so that the terms $\cos^2(\Delta m t/2)$ and $\sin^2
(\Delta m t/2)$ in (\ref{eqn:rat}) need not have the same coefficients, and a
$\cos \Delta m t$
term is generated in the rates.  This is the signal of ``direct'' CP violation,
as will be discussed below.  Its presence for $B \to \pi \pi$ was
pointed out by London and Peccei \cite{LPcos} and by Gronau.\cite{MGcos}

\subsection{Time-integrated asymmetry}

If one integrates the rates for $\bo_{t=0} \to f$ and $\ob_{t=0} \to \of$,
one can form the time-integrated asymmetry \cite{DR}
\beq
C_f \equiv
\frac{\Gamma(\bo_{t=0} \to f) - \Gamma(\ob_{t=0} \to \of)}
     {\Gamma(\bo_{t=0} \to f) + \Gamma(\ob_{t=0} \to \of)}~~~.
\eeq
If we consider the cases (as above) in which $|\mat{f}{\bo}| = |\mat{\of}
{\ob}|$, we just need the integral
\beq
\int_0^\infty \! dt \sin (\Delta m t) e^{- \Gamma t} = \frac{1}{\Gamma}
\frac{x_d}{1 + x_d^2}~~~,
\eeq
and we then find
\beq
C_f = - \frac{x_d}{1 + x_d^2} {\rm Im} \lambda_0
\eeq
when $|x| = 1$.  This is indeed maximal when $x_d = 1$; the
coefficient of $-{\rm Im} \lambda_0$ is 1/2.  For the actual value of
$x_d \simeq 0.75$, the coefficient is 0.48 instead, very close to its
maximal value.
 
\subsection{Specific examples in decays to CP eigenstates}

When $f$ is a CP eigenstate, a CP-violating difference between the rates for
$\bo \to f$ and $\ob \to \of$ arises as a result of interference between
the direct decays and those proceeding via mixing (i.e., $\bo \to \ob \to f$
and $\ob \to \bo \to \of$).  The second term in Eqs.\ (\ref{eqn:tdm})
is the result of this mixing.  As mentioned, the rate asymmetry goes to
zero when $x_d \to 0$ or $x_d \to \infty$.  We now illustrate the calculation
for two specific examples, $\bo \to J/\psi K_S$ and $\bo \to \pi \pi$.
\medskip

\leftline{\underline{The ``golden mode'':  $J/\psi K_S$}}

The quark subprocess governing $\bo \to J/\psi K_S$ is $\bar b \to \bar c c
\bar s$, whose CKM factor is $V^*_{cb} V_{cs}$.  The $K_S$ is produced through
its $\ko$ component.  The corresponding decay $\ob \to J/\psi K_S$ proceeds
via $b \to c \bar c s$ and involves the $\ok$ component of $K_S$.

For a CP-eigenstate, we defined $x \equiv \mat{f}{\ob}/\mat{f}{\bo}$ and
$\lambda_0 = (q/p)x$, but what we actually calculate is $\mat{\of}{\ob}/
\mat{f}{\bo}$ where $\of = \eta^f_{CP} f$ with $\eta^f_{CP} = \pm 1$.
For $f = J/\psi K_S$, $\eta^f_{CP} = -1$.  To show this, note that
$CP \ket{K_S} = \ket{K_S}$ and $CP \ket{J/\psi} = \ket{J/\psi}$ (since
$J/\psi$ has odd C and P).  The decay of the spin-zero $\bo$ to the spin-one
$J/\psi$ and the spin-zero $K_S$ produces the final particles in a state of
orbital angular momentum $\ell = 1$ and hence odd parity, introducing an
additional factor of $-1$.  Then
\beq
x = - \frac{\mat{K_S}{\ok}\mat{\ok}{\ob}}
           {\mat{K_S}{\ko}\mat{\ko}{\bo}}~~~.
\eeq
(A good discussion of the sign is given by Bigi and Sanda.\cite{BSsgn})
Now $\ket{K_S} = p_K \ket{\ko} + q_K \ket{\ok}$, so that $\mat{K_S}{\ok} =
q^*_K$ and $\mat{K_S}{\ko} = p^*_K$.  These numbers are very close to
$1/\s$.  If the loop calculation of $M_{12}$ for $\ko$--$\ok$ mixing
is dominated by the charmed quark, then $(q_K/p_K) \simeq (V_{cd} V^*_{cs})
/(V^*_{cd}V_{cs})$, and
\beq
\lambda_0 = - \frac{V^*_{cd}V_{cs}}{V_{cd}V^*_{cs}}
              \frac{V_{cb}V^*_{cs}}{V^*_{cb}V_{cs}}
              \frac{V_{td}V^*_{tb}}{V^*_{td}V_{tb}}~~~.
\eeq
We assumed a specific quark to dominate the calculation of
$M_{12}$ to illustrate the self-consistency of the expression
for $\lambda_0$ with with respect to redefinition of quark phases.  Note
first of all that the denominator is the complex conjugate of the numerator.
Then note that each quark is represented by the same number of $V$'s and
$V^*$'s in the numerator:  2 for the charmed quark and 1 each for $d,s,b$,
and $t$.  Thus any phase rotation of a quark field leaves the expression
invariant. (Bjorken and Dunietz have introduced a nice representation
of this invariance.\cite{BD})  The same cancellation would have occurred if
we were to say another quark dominated $\ko$--$\ok$ mixing.

For the final state $f= J/\psi K_S$ we thus find $\lambda_0 = - e^{- 2 i
\beta}$ and Im $\lambda_0 = \sin 2 \beta$, leading to the time-integrated
rate asymmetry $C_{J/\psi K_S} = -x_d \sin 2 \beta/(1+x_d^2)$.  In practice
the experiments often select events occurring for a proper time $t \ge t_0
> 0$ in order to enhance the signal/noise ratio, so that analyses are usually
based on the time-dependent asymmetry mentioned earlier.

%U                                  |||||||||
Some recent results on $\sin 2 \beta~^{58-62}$ are quoted in Table
\ref{tab:s2b}, and $\pm 1 \sigma$ limits from the average are plotted in Fig.\
\ref{fig:re}.  While the central value is somewhat below that favored by
%U                                                     
other observables, there is no significant discrepancy.
\medskip

\leftline{\underline{The $\pi^+ \pi^-$ mode and its complications}}

The main subprocess in $\bo \to \pi^+ \pi^-$ is the ``tree''
diagram in which $\bar b \to \pi^+ \bar u$, with the spectator $d$ combining
with the $\bar u$ to make a $\pi^-$.  Let us temporarily assume this is
the only important process and compute the CP-violating rate asymmetry.  We
shall return in Section 11 to the important role of ``penguin'' diagrams.

\begin{table}
\caption{Values of $\sin 2 \beta$ implied by recent measurements of the
CP-violating asymmetry in $\bo \to J/\psi K_S$. \label{tab:s2b}}
\begin{center}
\begin{tabular}{c c} \hline
\protect
Experiment & Value \\ \hline
OPAL \cite{OPs2b}  & $3.2^{+1.8}_{-2.0} \pm 0.5$ \\
CDF \cite{CDFs2b}  & $0.79^{+0.41}_{-0.44}$ \\
ALEPH \cite{ALs2b} & $0.84^{+0.82}_{-1.04} \pm 0.16$ \\
Belle \cite{Bes2b} & $0.58^{+0.32+0.09}_{-0.34-0.10}$ \\
BaBar \cite{Bas2b} & $0.34 \pm 0.20 \pm 0.05$ \\ \hline
%U                       |       ||
Average            & $0.48 \pm 0.16$ \\
\hline
\end{tabular}
\end{center}
\end{table}

Since the (spin-zero) $\pi^+ \pi^-$ system in $\bo$ decay has even CP, we find
\bea
x & \equiv & \frac{\mat{\pi^+ \pi^-}{\ob}}{\mat{\pi^+ \pi^-}{\bo}}
= \frac{V_{ub}V^*_{ud}}{V^*_{ub}V_{ud}}~~~,\\  
\lambda_0 = \frac{q}{p} x & = & \frac{V_{td}V^*_{tb}}{V^*_{td}V_{tb}}
\frac{V_{ub}V^*_{ud}}{V^*_{ub}V_{ud}} = e^{-2 i \beta} e^{-2 i \gamma}~~~.
\eea
{\it Exercise: Check the invariance of this expression under redefinitions
of quark phases.}

Since $\beta + \gamma = \pi - \alpha$, we have $\lambda_0 = e^{2 i \alpha}$,
Im($\lambda_0) = \sin 2 \alpha$, and $C_{\pi^+ \pi^-} = - x_d \sin 2 \alpha
/(1+x_d^2)$.  [Remember that our asymmetries are defined in terms of
$(\bo - \ob)/(\bo + \ob)$.]
This result is limited in its usefulness for several reasons.

(a) Our neglect of penguin diagrams will turn out to make a big difference.

(b) The range of $\sin \alpha$ is large enough that early asymmetry
measurements are unlikely to expose contradictions with the standard
prediction.

(c) An even larger range of negative $\sin 2 \alpha$ turns out to be allowed
if $V_{cb}$ is larger than assumed in Sec.\ 4.

An interesting exercise (whose result would, of course, be modified by
penguin contributions) is to suppose that the asymmetries in $\bo \to
J/\psi K_S$ and $\bo \to \pi^+ \pi^-$ are due {\it entirely} to mixing
(i.e., to a ``superweak'') interaction.\cite{WSW}  In this case, since
$J/\psi K_S$ and $\pi^+ \pi^-$ have opposite CP eigenvalues, one has
$C_{\pi^+ \pi^-} = - C_{J/\psi K_S}$.  What range of parameters in the
standard CKM picture would imitate this relation?  In other words, for
what $\rho$ and $\eta$ would one have $\sin 2 \alpha = - \sin 2 \beta$?
[The answer is $\eta = (1 - \rho)\sqrt{\rho/(2 - \rho)}$.]

\section{Decays to CP-noneigenstates}

If the final state $f$ is not a CP eigenstate, i.e. if $f \ne \pm \of$,
as in the case $f = K^+ \pi^-$, $\of = K^- \pi^+$, then a CP-violating
rate asymmetry requires two interfering decay channels with different weak
and strong phases:
\bea
A(B \to f) & = & A_1 e^{i \phi_1} e^{i \delta_1}
               + A_2 e^{i \phi_2} e^{i \delta_2}~~~,\\
A(\bar B \to \of) & = & A_1 e^{-i \phi_1} e^{i \delta_1}
                         + A_2 e^{-i \phi_2} e^{i \delta_2}~~~.
\eea
Here the weak phases $\phi_i$ change sign under CP conjugation, while the
strong phases $\delta_i$ do not.  Define $\Delta \phi = \phi_1 - \phi_2$,
$\Delta \delta = \delta_1 - \delta_2$, and
\beq
{\cal A}(f) \equiv \frac{|A(B\to f)|^2 - |A(\bar B \to \of)|^2}
                        {|A(B\to f)|^2 + |A(\bar B \to \of)|^2}~~~.
\eeq
Then
\beq \label{eqn:as}
{\cal A}(f) = \frac{- 2 A_1 A_2 \sin \Delta \phi \sin \Delta \delta}
{A_1^2 + A_2^2 + 2 A_1 A_2 \cos \Delta \phi \cos \Delta \delta}~~~.
\eeq

\subsection{Examples of interesting channels}

\medskip
\leftline{\underline{$\bo \to K^+ \pi^-$ vs.\ $\ob \to K^- \pi^+$}}

\begin{figure}
\centerline{\epsfysize=1.55in \epsffile{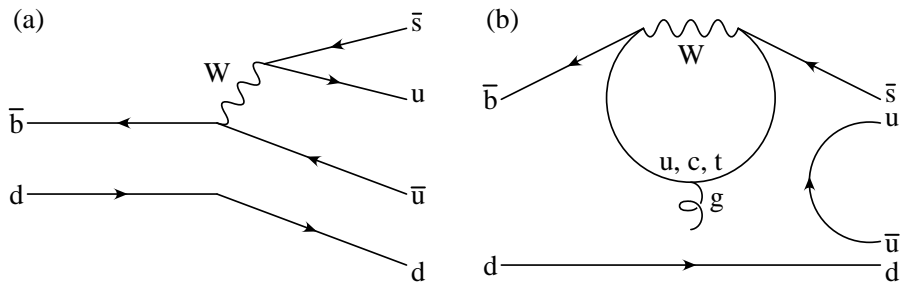}}
\caption{Contributions to $\bo \to K^+ \pi^-$.  (a) Color-favored
``tree'' amplitude
$\sim V^*_{ub}V_{us}$; (b) ``penguin'' amplitude $\sim V^*_{tb}V_{ts}$.}
\label{fig:kpi1}
\end{figure}

We illustrate two types of contribution to $\bo \to K^+ \pi^-$ in
Fig.\ \ref{fig:kpi1}.  The ``tree'' contribution, which in this case is
color-favored since the color-singlet current can produce a quark pair
of any color, has weak phase $\gamma = {\rm Arg}(V^*_{ub}V_{us})$ and strong
phase $\delta_T$, while the ``penguin'' contribution has weak phase
$\pi = {\rm Arg}(V^*_{tb}V_{ts})$ and strong phase $\delta_P$. 

% This is Figure 3

Even though $\delta_T - \delta_P$ is unknown, and may be small so that
little CP-violating asymmetry is present in $B \to K^\pm \pi^\mp$, it will turn
out that one can use rate information for several processes, with the help of
flavor SU(3) (which can be tested) to learn weak phases such as $\gamma$.
\medskip

\leftline{\underline{$B^+ \to K^+ \pi^0$ vs.\ $B^- \to K^- \pi^0$}}

{\it Exercise:  Identify the main amplitudes which contribute.  What are the
differences with respect to $B \to K^\pm \pi^\mp$?}

\begin{figure}
\centerline{\epsfysize=1.6in \epsffile{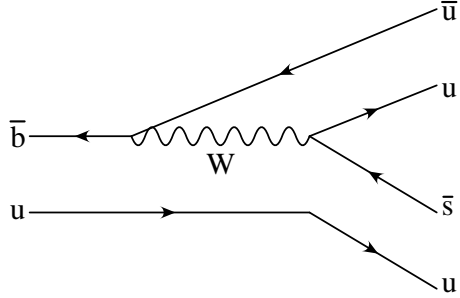}}
\caption{Color-suppressed tree diagram contributing to $B^+ \to K^+ \pi^0$.}
\label{fig:kpi2}
\end{figure}

Answer:  There are {\it two} ``tree'' amplitudes, one color-favored [as in
Fig.\ 3(a)] and one
color-suppressed (Fig.\ \ref{fig:kpi2}).  Both have weak phases
$\gamma = {\rm Arg}(V^*_{ub}V_{us})$.  There is a penguin amplitude [as in
Fig.\ 3(b)] with
weak phase $\pi = {\rm Arg}(V^*_{tb}V_{ts})$.  Since $\pi^0 = (d \bar d
- u \bar u)/\s$ in a phase convention in which $\pi^+ = u \bar d$, the
color-favored tree and penguin amplitudes are the same as that in
$\bo \to K^+ \pi^-$, but divided by $\s$.  Thus the overall rate for
$B^\pm \to K^\pm \pi^0$ is expected to be 1/2 that for $B \to K^\pm \pi^\mp$
if the penguin amplitude dominates or if the color-suppressed amplitude is
negligible.  In that case one expects similar CP-violating asymmetries
for $\bo \to K^+ \pi^-$ and $B^+ \to K^+ \pi^0$. \cite{GRcomb,MNKpi}
\medskip

% This is Figure 4

\medskip
\leftline{\underline{$B^+ \to \ko \pi^+$ vs.\ $B^- \to \ok \pi^-$}}

{\it Exercise:  Show that there is no tree amplitude and hence no CP-violating
asymmetry expected.}  This process is expected to be dominated by the
penguin amplitude and thus provides a reference for comparison with other
processes in which tree amplitudes participate.
\medskip

Small contributions to $B^+ \to K^+ \pi^0$ and $B^+ \to K^0 \pi^+$ are
possible from the process in which the $\bar b u$ pair annihilates into a
weak current which then produces $\bar s u$.  A $q \bar q$ pair is produced
in hadronization, giving $K^+ \pi^0$ if $q = u$ and $K^0 \pi^+$ if $q = d$.
These contributions are expected to be suppressed by a factor of $f_B/m_B$
if the graphs describing them can be taken literally.  However, they can also
be generated by rescattering from other contributions, e.g., $(B^+ \to K^+
\pi^0)_{\rm tree} \to K^0 \pi^+$.  We shall mention tests for such
effects in Sec.~12.

\subsection{Pocket guide to direct CP asymmetries}

We now indicate a necessary (but not sufficient) condition for the
observability of direct CP asymmetries based on the interference of two
amplitudes, one weaker than the other.  The result is that one must
be able to detect processes at the level of the {\it absolute square of the
weaker amplitude.}\cite{EGR} This guides the choice of processes in which one
might hope to see direct CP-violating rate asymmetries.

Suppose the weak phase difference $\Delta \phi$ and the strong phase difference
$\Delta \delta$ are both near $\pm \pi/2$ (the most favorable case for
detection of an asymmetry).  Then the asymmetry ${\cal A}$ in Eq.\
(\ref{eqn:as}) has magnitude
\beq
|{\cal A}| = {\cal O} \left( \frac{2 A_1 A_2}{A_1^2 + A_2^2} \right) \simeq
\frac{2 A_2}{A_1}~~~{\rm for}~A_2 \ll A_1~~~.
\eeq
Imagine a rate based on the square of each amplitude:  $N_i = {\rm const.}~
|A_i|^2$.  Then $|{\cal A}| \simeq 2 \sqrt{N_2/N_1}$.

The statistical error in ${\cal A}$ is based on the total number of events.
For $A_2 \ll A_1$, one has $\delta {\cal A} \simeq 1/\sqrt{N_1}$.  Then the
significance of the asymmetry (in number of standard deviations) is
\beq
\frac{{\cal A}}{\delta{\cal A}} \sim {\cal O}(2 \sqrt{N_2})~~~.
\eeq
Thus (aside from the factor of 2) one must be able to see the {\it square of
the weaker amplitude} at a significant level in order to see a significant
asymmetry due to $A_1$--$A_2$ interference.

\subsection{Interesting levels for charmless $B$ decays}

Typical branching ratios for the dominant $B$ decays to pairs of light
pseudoscalar mesons are in the range of 1 to 2 parts in $10^5$.
Some recent data are summarized in Table \ref{tab:PP}.
Here the average between a process and its charge conjugate is quoted.
%U Updated references
These data are based on results by CLEO,\cite{CLEOkpi,CLetap,CL2K} [including
a value for ${\cal B}(B^+ \to \pi^+ \pi^0)$ extracted from an earlier CLEO
report \cite{GRVP}], Belle,\cite{BePP} and BaBar.\cite{BaPP} The averages
are my own.

%U Updated table
\renewcommand{\arraystretch}{1.2}
\begin{table}[h]
\caption{Branching ratios, in units of $10^{-6}$ for $\bo$ or $B^+$ decays
to pairs of light pseudoscalar mesons. \label{tab:PP}}
\begin{center}
\begin{tabular}{c c c c c} \hline
Mode          & CLEO \cite{CLEOkpi,CLetap,CL2K,GRVP}
                & Belle \cite{BePP}
                  & BaBar \cite{BaPP}
                    & Average \\
$\pi^+ \pi^-$ & $4.3^{+1.6}_{-1.4} \pm 0.5$
                & $5.6^{+2.3}_{-2.0} \pm 0.4$
                  & $4.1 \pm 1.0 \pm 0.7$
                    & $4.4 \pm 0.9$ \\
$\pi^+ \pi^0$ & $5.4 \pm 2.6$
                & $7.8^{+3.8+0.8}_{-3.2-1.2}$
                  & $5.1^{+2.0}_{-1.8} \pm 0.8$
                    & $5.6 \pm 1.5$ \\
$K^+ \pi^-$   & $17.2^{+2.5}_{-2.4} \pm 1.2$
                & $19.3^{+3.4+1.5}_{-3.2-0.6}$
                  & $16.7 \pm 1.6^{+1.2}_{-1.7}$
                    & $17.4 \pm 1.5$ \\
$\ko \pi^+$   & $18.2^{+4.6}_{-4.0} \pm 1.6$
                & $13.7^{+5.7+1.9}_{-4.8-1.8}$
                  & $18.2^{+3.3+1.6}_{-3.0-2.0}$
                    & $17.3 \pm 2.4$ \\
$K^+ \pi^0$   & $11.6^{+3.0+1.4}_{-2.7-1.3}$
                & $16.3^{+3.5+1.6}_{-3.3-1.8}$
                  & $10.8^{+2.1+1.0}_{-1.9-1.2}$
                    & $12.2 \pm 1.7$ \\
$\ko \pi^0$   & $14.6^{+5.9+2.4}_{-5.1-3.3}$
                & $16.0^{+7.2+2.5}_{-5.9-2.7}$
                  & $8.2^{+3.1+1.1}_{-2.7-1.2}$
                    & $10.4 \pm 2.6$ \\
$K^+ \eta'$ & $80^{+10}_{-9} \pm 7$ & & $62 \pm 18 \pm 8$ & $75 \pm 10$ \\
$\ko \eta'$ & $89^{+18}_{-16} \pm 9$ & & & $78 \pm 9$ (a) \\ \hline
\end{tabular}
\end{center}
\leftline{(a) Average for $K^+ \eta'$ and $K^0 \eta'$ modes.}
\end{table}
 
The relative $K \pi$ rates are compatible with dominance by the penguin
amplitude, which predicts the rates involving a neutral pion to be half those
with a charged pion.
This conclusion is supported by an estimate of the tree contribution via
the decay $B \to \pi \ell \nu$ and factorization.  One then needs some idea
of the form factor at $m(\ell \nu) = m_\pi$ or $m_K$.  The result is that
one estimates ${\cal B}_{\rm tree}(\bo \to \pi^+ \pi^-) \simeq 10^{-5}$, or
\beq
{\cal B}_{\rm tree}(\bo \to K^+ \pi^-) \simeq \left( \frac{f_K}{f_\pi}
\right)^2 \left| \frac{V_{us}}{V_{ud}} \right|^2 \times 10^{-5}~~~.
\eeq
With $f_K = 161$ MeV, $f_\pi = 132$ MeV, $f_K/f_\pi = 1.22$, $V_{us}/V_{ud}
= \tan \theta_c = 0.22/0.975 = 0.226$, the coefficient of $10^{-5}$ on the
right-hand side is 0.076.  Thus in order to see a significant CP-violating
rate asymmetry in $B \to K \pi$ one needs at least 13 times the sensitivity
that was needed in order to see all the $B \to K \pi$ modes.  This would
correspond to about 100 fb$^{-1}$ at $e^+ e^-$ colliders, or samples of about
$10^8$ identified $B$'s at hadron machines.  In other words, one needs to be
able to see branching ratios of  a few parts in $10^7$ with good statistical
significance.  This is within the capabilities of experiments just now
getting under way.

\section{Flavor tagging}

\subsection{States of $B \bar B$ with definite charge-conjugation}

The process $e^+ e^- \to B \bar B$ is typically studied at the mass of the
$\Upsilon(4S)$ resonance, $E_{\rm c.m.} = 10.58$ GeV, above the threshold of
$2 M_B = 10.56$ GeV for $B \bar B$ production but below the threshold for
production of one or two $B^*$'s:  $M_B + M_{B^*} = 10.605$ GeV, $2 M_{B^*}
= 10.65$ GeV.  Now, the $\Upsilon(4S)$ has $C = -1$.  It
is produced via a virtual photon $\gamma^*$ (which has odd $C$).  
It is a $^3S_1$ $b \bar b$ state, where the superscript $2 S_{b \bar b} + 
1$ denotes the total spin $S_{b \bar b} = 1$.  The $b \bar b$ pair has orbital
angular momentum $L = 0$ and total angular momentum $J=1$.  A $Q \bar Q$ state
$^{2S+1}L_J$ in general has $C = (-1)^{L+S}$.

The $B \bar B$ pair produced at the $\Upsilon(4S)$ thus has a definite
eigenvalue of charge-conjugation, $C(B \bar B) = -1$, correlating the flavor
flavor of the neutral $B$ whose decay (e.g., to $J/\psi K_S$) is being studied
with the flavor of the other $B$ used to ``tag'' the decay, e.g., via a
semileptonic decay $b \to c \ell^- \bar \nu_\ell$ or $\bar b \to \bar c \ell^+
\nu_\ell$.

Let $\bo \ob$ be in an eigenstate of C with eigenvalue $\eta_C = \pm 1$.  (To
get a state with $\eta_C = +1$ it is sufficient to utilize the reaction
$e^+ e^- \to {\bo}^* \ob~{\rm or}~\bo {\ob}^* \to \bo \ob \gamma$ just above
threshold.)  In the $\bo \ob$ center-of-mass system, the wave function of the
pair,
\beq
\Psi_C \equiv \frac{1}{\s} \left[ \bo(\hp) \ob(-\hp) + \eta_C \ob(\hp)
\bo(-\hp) \right]
\eeq
may be expressed in terms of the mass eigenstates $B_{L,H}$ in order to
study its time evolution.  Since
\beq
\bo = \frac{1}{p \s} \left[ B_L + B_H \right]~~,~~~
\ob = \frac{1}{q \s} \left[ B_L - B_H \right]~~~,
\eeq
we have
$$
 \Psi_C \equiv \frac{1}{\s} \frac{1}{2 p q} 
  \left\{ [B_L(\hp) + B_H(\hp)][B_L(-\hp) - B_H(-\hp)] \right.
$$
\beq
 \left. + \eta_C [B_L(\hp) -B_H(\hp)][B_L(-\hp) +  B_H(-\hp)] \right\}~~~.
\eeq

For $\eta_C = -1$ the $LL$ and $HH$ terms cancel (this is also a consequence
of Bose statistics) and one has
\beq \label{eqn:ecm}
\Psi_C(\eta_C = -1) = \frac{1}{\s p q}
  \left[ B_H(\hp) B_L(-\hp) - B_L(\hp) B_H(-\hp) \right]~~~.
\eeq
For $\eta_C = + 1$ the $HL$ and $LH$ terms cancel and one has
\beq \label{eqn:ecp}
\Psi_C(\eta_C = +1) = \frac{1}{\s p q} 
  \left[ B_L(\hp) B_L(-\hp) - B_H(\hp) B_H(-\hp) \right]~~~.
\eeq
Define $t$ and $\ot$ to be the proper times with which the states $\hp$
and $-\hp$ evolve, respectively:
\beq
B_{L,H}(\hp) \to B_{L,H}(\hp)e^{-i \mu_{L,H} t}~~,~~~
B_{L,H}(-\hp) \to B_{L,H}(-\hp)e^{-i \mu_{L,H} \ot}~~~.
\eeq
Project the state with $\hp$ into the desired decay mode (e.g., $J/\psi K_S$)
and the state with $-\hp$ into the tagging mode (which signifies $\bo$ or
$\ob$ at time $\ot$, e.g., $\ell^- \leftrightarrow \ob,~\ell^+ \leftrightarrow
\bo$).  Then, for a CP-eigenstate, it is left as an {\it Exercise} to show
in the limit $\Delta \Gamma = 0$ that
\bea \label{eqn:ttb}
\left. \frac{d^2 \Gamma[f(t),\ell^-(\ot)]}{d t~ d \ot} \right|_{\eta_C = \mp 1}
& \sim & e^{- \Gamma(t + \ot)} [ 1 - \sin \Delta m (t \mp \ot) {\rm Im}
\lambda]~~,\\
\left. \frac{d^2 \Gamma[f(t),\ell^+(\ot)]}{d t~ d \ot} \right|_{\eta_C = \mp 1}
& \sim & e^{- \Gamma(t + \ot)} [ 1 + \sin \Delta m (t \mp \ot) {\rm Im}
\lambda]~~~.
\eea
(Hints:  Recall that $\lambda = (q/p)\mat{f}{\ob}/ \mat{f}{\bo}$, $\bar
\lambda = (p/q)(\mat{\of}{\bo}/ \mat{\of}{\ob}$, $|\lambda| = 1$, $\bar \lambda
= \lambda^*$, $\Delta m = m_H - m_L$.  For $\eta_C = -1$, write
the decay amplitude as a function of $t$ and $\ot$. It will have two terms,
one $\sim e^{-i(m_H t + m_L \ot)}$ and the other $\sim e^{-i(m_L t + m_H 
\ot)}$, whose interference in the absolute square of the amplitude
gives rise to the $\sin \Delta m(t - \ot)$ terms.)
These results have some notable properties.

(1) For either
value of $\eta_C$, the sum of the $\ell^+$ and $\ell^-$ results is as if one
didn't tag, and the oscillatory terms cancel one another.

(2) For $\eta_C =
-1$, note the {\it antisymmetry} with respect to $t - \ot$.  This is a
consequence of the Bose statistics and the $C = -1$ nature of the initial
state.  If one integrates over all times, the CP-violating asymmetry vanishes.
Thus in order for the tagging method to work in a C-odd state like
$\Upsilon(4S)$ one must know whether $t$ or $\ot$ was earlier.  An asymmetric
$B$-factory like PEP-II or KEK-B permits this by spreading out the decay using
a Lorentz boost.

{\it Exercise:  Show for $\eta_C = -1$ that if one subdivides the $t,\ot$
integrations according to $t < \ot$ or $t > \ot$, then}
$$
\frac{\int \! \int dt d \ot (d^2 \Gamma/d t d \ot) [(\ell^-, t > \ot)
- (\ell^-, t < \ot) - (\ell^+, t > \ot) + (\ell^+, t < \ot)]}
{\int \! \int dt d \ot (d^2 \Gamma/d t d \ot) [(\ell^-, t > \ot)
+ (\ell^-, t < \ot) + (\ell^+, t > \ot) + (\ell^+, t < \ot)]}
$$
\beq
= - \frac{x_d}{1 + x_d^2} {\rm Im} \lambda~~~.
\eeq
In practice the BaBar and Belle analyses will probably fit the time
distributions rather than simply subdividing them, since background rejection
and signal/noise ratio are functions of $t-\ot$.

(3) For $\eta_C = +1$ the oscillatory term behaves as $\sim \Delta m (t +
\ot)$, so it is not necessary to know whether $t$ or $\ot$ was earlier,
and the asymmetric collision geometry is not needed.  However, as shown above,
in order to produce a $\bo \ob$ state with $\eta_C = +1$ in $e^+ e^-$
collisions one must work at or above $B \bar B^*$ threshold, thereby losing
the cross section advantage of the $\Upsilon(4S)$ resonance.
 
\subsection{Uncorrelated $B \bar B$ pairs}

Pairs of $B$'s produced in a hadronic environment are likely to arise
from independent fragmentation of $b$ and $\bar b$ quarks, so that it is
unlikely that they are produced in a state of definite $\eta_C$.  (The
interesting case of partially-correlated $B$-$\bar B$ pairs can be
attacked by density-matrix methods.\cite{GRdens})  Thus, one must resort
to either the fact that a $b$ is always produced in association with a $\bar
b$ by the strong interactions (``opposite-side tagging''), or the fact that
the fragmentation of a $b$ into a $\ob$ favors one particular sign of charged
pion close to the $\ob$ in phase space (``same-side tagging'').
\medskip

\leftline{\underline{``Opposite-side'' methods}}

The strong interactions $q \bar q \to b \bar b$ or $g g \to b \bar b$
($g = $ gluon) conserve beauty, so that a $b$ can be identified if it is
found to be produced in association with a $\bar b$.  The opposite-side
$\bar b$ can be identified in several ways.

(1) The {\it jet-charge} method makes use of the fact that a jet tends to
carry the charge of its leading quark \cite{FF}, since the average charge
of the fragmentation products is zero in the flavor-SU(3) limit.
(There is some delicacy if strange
quark production is suppressed, since $Q(u) \ne - Q(d)$.\cite{FR})

(2) The {\it lepton-tag} method uses the charge of the lepton in the
semileptonic decays $b \to c \ell^- \bar \nu_\ell$ and $\bar b \to \bar c
\ell^+ \nu_\ell$ to signify the flavor of the decaying opposite-side $b$
quark.  The signal is diluted since semileptonic decays occur after the $b$
quark has been incorporated into a meson.  Sometimes this meson is a
neutral $B$, in which case information on its flavor is nearly completely
lost if it is a strange $B$ and partially lost if it is a nonstrange $B$.
An initial $\overline{B}_s$ will decay half the time as a $B_s$, while
an initial $\ob$ will decay about 18\% of the time as a $\bo$.  (See Section
5.2.)

(3) The {\it kaon-tag} method uses the fact that the signs of kaons
produced in $b$ decays are correlated with the flavor of the decaying
$b$.  A $b$ gives rise to $c$, whose products have more $K^-$ than $K^+$.
This method is subject to the same dilution as the lepton-tag method.

In order to utilize the above methods, one needs the relative
probabilities of production of $B^0$, $B^+$, $B_s$, and $\Lambda_b$,
The CDF Collaboration \cite{CDFbpr} has measured these in high-energy hadron
collisions to be in the ratios 0.375:0.375:0.16:0.09, while LEP Collaborations
find 0.40:0.40:0.097:0.104 for $Z^0 \to b \bar b$ decays.\cite{LEPbpr}
Taking account of $P(\bo \to \ob) \simeq
18\%$ and $P(B_s \to \overline{B}_s) \simeq 1/2$, the probability of a
``wrong'' tag is $(3/8)(0.18) + (0.16)(1/2) \simeq 0.15$, which dilutes the
efficacy of the tag by a factor (right -- wrong)/(right + wrong) $\simeq 0.70$.

For an extensive study of the first two tagging methods, see recent papers by
the CDF Collaboration.\cite{CDFtags}  These methods were a key ingredient in
obtaining the CP-violating asymmetry in $\bo \to J/\psi K_S$ mentioned in
Sec.~6.4.
\medskip

\leftline{\underline{``Same-side'' methods:  Fragmentation and $B^{**}$
resonances}}

The fragmentation of a $b$ quark into a neutral $B$ meson
is not charge-symmetric.
This was noted quite some time ago in the context of strange $B$'s.\cite{Bstag}
A $B_s$ contains a $\bar b$ and an $s$.  This $s$ must have been produced in
association with a $\bar s$.  If that $\bar s$ is incorporated into a charged
kaon, the kaon must be a $K^+ = u \bar s$.

A similar argument applies to non-strange neutral $B$'s and charged
pions.\cite{GNR} A $\bo$ is then found to be associated more frequently with a
$\pi^+$ nearby in phase space, while a $\ob$ tends to be associated with a
$\pi^-$.  This correlation is the same as that found in resonance decays:
$\bo$ resonates with $\pi^+$ but not $\pi^-$, while $\ob$ resonates with
$\pi^-$ but not $\pi^+$.
\medskip

\begin{figure}
\centerline{\epsfysize=1.5in \epsffile{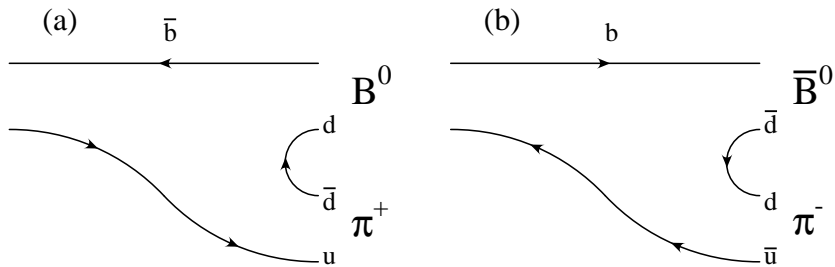}}
\caption{Fragmentation of a $\bar b$ or $b$ quark into a $\bo$ or $\ob$.}
\label{fig:frag}
\end{figure}

The fragmentation of a $\bar b$ or $b$ quark is illustrated in Fig.\
\ref{fig:frag}.  If one cuts the diagrams to the left of the pion emission,
one finds either a $\bar b u$ or a $b \bar u$ state.  Thus, a positively
charged resonance can decay to $\bo \pi^+$, while a negatively charged one
can decay to $\ob \pi^-$. 

% This is Figure 5

Recall the case of $D^{*+} \to \pi^+ D^0$ mentioned in Sec.~3.1.  The soft
pion in that decay may be used to tag the flavor of the neutral $D$ at the
time of its production, which is useful if one wants to study $\dz$--$\od$
mixing or Cabibbo-disfavored decays.  The difference between $M(D^{*+})$ and
$M(\pi^+) + M(\dz)$ (the ``$Q$-value'') is only about 5 MeV, so the pion is
nearly at rest in the $D^{*+}$ c.m.s., and hence kinematically very
distinctive.  In the case of $B$'s, however, the lowest vector meson $B^{*+}$
cannot decay to $\pi^+ \bo$ since the $B^*$--$B$ mass difference is only
about 46 MeV.  One must then utilize the decays of higher-mass $B^*$'s,
collectively known as $B^{**}$'s.  The lightest such states consist of
a $b$ quark and a light quark ($u,d,s$) in a P-wave.

Many key features of the spectroscopy of a heavy quark and a light one were
first pointed out in the case of charm,\cite{DGG,JRPW} and codified using
heavy-quark symmetry.\cite{HQSPW}  The heavy quark spin degrees of freedom
nearly decouple from the light-quark and gluon dynamics, so it makes sense
to first couple the relative angular momentum $L = 1$ and the light-quark
spin $s_q = 1/2$ to states of total light-quark angular momentum $j_q =
1/2$ or 3/2 and then to couple $j_q$ with the heavy quark spin $S_{\bar Q} =
1/2$ to form total angular momentum $J$.  For $j_q = 1/2$ one then gets
states with $J=0,1$, while for $j_q = 3/2$ one gets states with $J=1,2$.
The $j_q = 1/2$ states decay to $B \pi$ or $B^* \pi$ only via S-waves, while
the $j_q = 3/2$ states decay to $B \pi$ or $B^* \pi$ only via D-waves.  These
properties are summarized in Table \ref{tab:PWB}.

\renewcommand{\arraystretch}{1.0}
\begin{table}
\caption{Quantum numbers of $B^{**}$ resonances (P-wave resonances between
a $\bar b$ quark and a light quark $q$). \label{tab:PWB}}
\begin{center}
\begin{tabular}{c c c c c} \hline
$j_q$ & $J$ & Decay prods.       & Part.\ wave & Width \\ \hline
 1/2  &  0  & $B \pi$            & S wave      & Broad \\
 1/2  &  1  & $B^* \pi$          & S wave      & Broad \\
 3/2  &  1  & $B^* \pi$          & D wave      & Narrow \\
 3/2  &  2  & $B \pi$, $B^* \pi$ & D wave      & Narrow \\ \hline
\end{tabular}
\end{center}
\end{table}

The $j_q=3/2$ resonances, decaying via D-waves,
have been seen, with typical widths of tens of MeV and masses somewhere
between 5.7 and 5.8 GeV/$c^2$ in all analyses.\cite{CDFB**,LEPB**} The $j_q =
1/2$ resonances are expected to be considerably broader.  There is no
unanimity on their properties, but evidence exists for at least one of their
charmed counterparts.\cite{CLEOD**} More information on $B^{**}$'s
would enhance their usefulness in same-side tagging of neutral $B$ mesons.

\section{The strange $B$}

\subsection{$B_s$--$\overline{B}_s$ mixing}

The limit \cite{Bslim} $\Delta m_s > 15$ ps$^{-1}$ mentioned in Sec.\ 4.4 was
a significant source of constraint on the $(\rho,\eta)$ plot.  What range of
mixing is actually expected?  One may place an upper bound by noting that
\beq
\frac{\Delta m_s}{\Delta m_d} = \left| \frac{V_{ts}}{V_{td}} \right|^2
\frac{B_{B_s}}{B_B} \left( \frac{f_{B_s}}{f_B} \right)^2~~~.
\eeq
Let us review the estimate \cite{JRFM} $f_{B_s}/f_B \le 1.25$.  The upper
limit (larger than the lattice range \cite{DeG}) comes from the nonrelativistic
quark model, which implies \cite{NRFM}
$|f_M|^2 = 12 |\Psi(0)|^2/M_M$ for the decay constant $f_M$ of
a meson $M$ of mass $M_M$ composed of a quark-antiquark pair with relative
wave function $\Psi(\vec{r})$.  One estimates the ratios of $|\Psi(0)|^2$
in $D$ and $D_s$ systems from strong hyperfine splittings.  Since $M(D^{*+})
- M(D^+) \simeq M(D_s^{*+}) - M(D_s^+)$, one expects $|\Psi(0)|_{Q \bar d}^2
/m_d \simeq |\Psi(0)|_{Q \bar s}^2/m_s$ for mesons containing a heavy quark
$Q$.  In constituent-quark models \cite{GasR} $m_d/m_s \simeq
0.64$, so $f_{Q \bar d}/f_{Q \bar s} \simeq \sqrt{0.64} = 0.8$.
An upper limit on $|V_{ts}/V_{td}|$ (see Sec.\ 4.3) is then $\le
[\lambda (0.66)]^{-1} = 6.9$, implying $\Delta m_s/\Delta m_d \le 74$ or
$\Delta m_s \le 36$ ps$^{-1}$.  A recent prediction \cite{BecR} based on
lattice estimates for decay constants is $\Delta m_s = 16.2 \pm 2.1 \pm
3.4$ ps$^{-1}$; there is a hint of a signal at $\sim 17$ ps$^{-1}$.\cite{Bslim}

\subsection{$B_s$ lifetime}

The mass eigenstates of the strange $B$ are expected to be nearly CP
eigenstates.  Their mass splitting is expected to be correlated with their
mixing; large values of $\Delta m_s$ imply
large values of $\Delta \Gamma_s$.  (In the calculation of $\Delta \Gamma/\
\Delta m$, the values of $|f_M|^2$ cancel.)

\begin{figure}
\centerline{\epsfysize=1.4in \epsffile{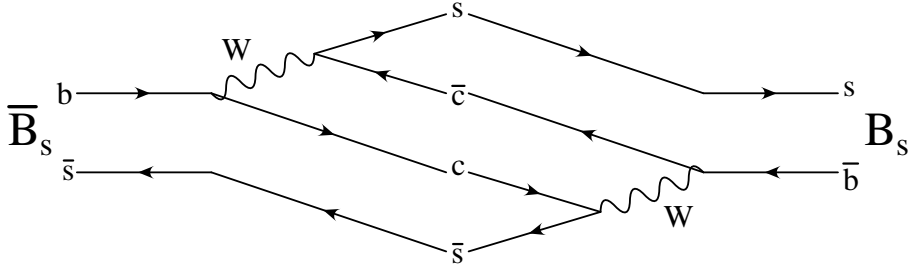}}
\caption{Mixing of $\overline{B}_s$ and $B_s$ as a result of shared
$s \bar c c \bar s$ intermediate states.}
\label{fig:bsmix}
\end{figure}

The value of $\Delta \Gamma_s$
is much greater than that of $\Delta \Gamma_d$ because of the shared
intermediate states in the transition $\overline{B}_s = b \bar s \to
s \bar c c \bar s \to s \bar b = B_s$ illustrated in Fig.\ \ref{fig:bsmix}.
Specific calculations \cite{dbs} imply that this quark subprocess may be
dominated by $CP = +$ intermediate states, implying a shorter lifetime for
the even-CP eigenstate of the $\overline{B}_s$--$B_s$ system by ${\cal O}
(10\%)$.

% This is Figure 6

The imaginary part of the diagram in Fig.\ \ref{fig:bsmix}
is due to on-shell states, which do not include those involving
the top quark.  This is the reason for the factor of $(m_b^2/m_W^2)/
S(m_t^2/M_W^2)$ in Eq.\ (\ref{eqn:bgm}).  The lowest-order ratio,
$\Delta \Gamma_s/\Delta m_s \simeq -1/180$, implies that if $|\Delta m_s/
\Gamma_s| = (20 {\rm~ps}^{-1})(1.6 {\rm~ps}) = 32$ (a reasonable value), then
$|\Delta \Gamma_s / \Gamma_s| \simeq 1/6$.  Recent calculations predict
$\Delta \Gamma_s / \Gamma_s = (9.3^{+3.4}_{-4.6})\%$ \cite{BeLe} or
$(4.7 \pm 1.5 \pm 1.6)\%$ \cite{bec}. The latter group \cite{beclat} also finds
$f_{B_d} \sqrt{B_{B_d}} = 206(28)(7)$ MeV, $f_{B_s} \sqrt{B_{B_s}}/(f_B
\sqrt{B_B}) = 1.16(7)$, $f_{B_s} \sqrt{B_{B_s}} = 237(18)(8)$ MeV.

\subsection{Measuring $\Delta \Gamma_s/ \Gamma_s$}

The average decay rate of the two mass eigenstates $B_H$ and $B_L$ can be
measured by observing a flavor-specific decay, e.g.,
\beq
B_s (= \bar b s) \to D_s^-(= \bar c s) \ell^+ \nu_\ell~~{\rm or}~~
B_s \to D_s^- \pi^+~~,~~~D_s^- \to \phi \pi^-~~~.
\eeq
The flavor of $D_s$ labels the flavor of the $B_s$ and then we note that
\beq
\ket{B_s} \simeq \frac{1}{\s} \left( \ket{B_L} + \ket{B_H} \right)~~~.
\eeq
Such a flavor-specific decay then gives a rate $\bar \Gamma = (\Gamma_L
+ \Gamma_H)/2$.

One can also look for a decay in which the CP of the final state can be
easily identified.\cite{DDLR}  $B_s \to J/\psi \phi$ is such a final state;
one can perform a helicity analysis to learn its CP eigenvalue (or whether
it is a mixture).  Since $J(J/\psi) = J(\phi) = 1$ and $J(B_s) = 0$, the final
state can have orbital angular momenta $\ell = 0, 1$, and 2.  {\it Exercise:
Show that $\ell = 0,2$ corresponds to even, and $\ell = 1$ to odd CP.}

\subsection{Helicity analyses of $B_s \to J/\psi \phi$ and $B \to J/\psi
K^{*}$}

It is convenient to re-express the three partial wave amplitudes for
decay of a spin-zero mesons into two massive spin-1 mesons in terms of {\it
transversity} amplitudes.\cite{transv}  These are most easily visualized by
analogy with the method originally used to determine the parity of the
neutral pion through its decay to two photons.

A spinless meson $M$ can decay to two photons with two possible linear
polarization states:  parallel and perpendicular to each other.  If they have
parallel polarizations, the interaction Lagrangian is ${\cal L}_{\rm int} \sim
M F_{\mu \nu} F^{\mu \nu} \sim M({\bf E}^2 - {\bf B}^2)$, while if they
have perpendicular polarizations, ${\cal L}_{\rm int} \sim M F_{\mu \nu}
\tilde{F}^{\mu \nu} \sim M({\bf E} \cdot {\bf B})$.  Now, ${\bf E}^2 -
{\bf B}$ is CP-even, while ${\bf E} \cdot {\bf B}$ is CP-odd.  The observation
that the two photons emitted by the $\pi^0$ had perpendicular polarizations
then was used to infer that the pion had odd CP and hence (since its C was
even as a result of its coupling to two photons) odd P.

One can then identify two of the decay amplitudes for a spinless meson
decaying to two {\it massive} vector mesons as $A_\pr$ (parallel linear
polarizations, even CP) and $A_\perp$ (perpendicular linear polarizations,
odd CP).  A third decay amplitude is peculiar to the massive vector meson
case:  Both vector mesons can have longitudinal polarizations (impossible
for photons).  Since there must be {\it two} independent CP-even decay
amplitudes by the partial-wave exercise given above, this amplitude,
which we call $A_0$, must be CP-even.

There are two recent experimental studies of decays of strange $B$'s to
pairs of vector mesons. (1)  The CDF Collaboration \cite{CDFBs} finds the
results quoted in Table \ref{tab:Bhel}.  The decays $B_s \to J/\psi \phi$
and $\bo \to J/\psi K^{*0}$ are related to one another by flavor SU(3) (the
interchange $s \leftrightarrow d$ for the spectator quark) and thus should
have similar amplitude structure.  We have adopted a normalization in
which $|A_0|^2 + |A_\pr|^2 + |A_\perp|^2 = 1$.  The CDF result says that
$B_s \to J/\psi \phi$ is dominantly CP-even.  No significant $\Delta \Gamma
/\Gamma$ has been detected, but the sensitivity is not yet adequate to reach
predicted levels.  These conclusions are supported by results from CLEO
\cite{CLhel} ($|A_0|^2 = 0.52 \pm 0.08,~|A_\perp|^2 = 0.09 \pm 0.08$) and
BaBar \cite{Bahel} ($|A_0|^2 = 0.60 \pm 0.06 \pm 0.04,~|A_\perp|^2 = 0.13
\pm 0.06 \pm 0.02$).

\begin{table}[h]
\caption{Amplitudes in the decays $B_s \to J/\psi \phi$ and $\bo \to J/\psi
K^{*0}$. \label{tab:Bhel}}
\begin{center}
\begin{tabular}{c c c} \hline
Amplitudes & $B_s \to J/\psi \phi$ & $\bo \to J/\psi K^{*0}$ \\ \hline
$|A_0|$     & $0.78 \pm 0.09 \pm 0.01$ & $0.77 \pm 0.04 \pm 0.01$ \\ 
$|A_\pr|$  & $0.41 \pm 0.23 \pm 0.05$ & $0.53 \pm 0.11 \pm 0.04$ \\
Arg($A_\pr/A_0$) & $1.1 \pm 1.3 \pm 0.2$ & $2.2 \pm 0.5 \pm 0.1$ \\
$|A_\perp|$ & $0.48 \pm 0.20 \pm 0.04$ & $0.36 \pm 0.16 \pm 0.08$ \\
Arg($A_\perp/A_0$) &                   & $-0.6 \pm 0.5 \pm 0.1$ \\ \hline
$\Gamma_L/\Gamma = |A_0|^2$ & $0.61 \pm 0.14 \pm 0.02$ & $0.59 \pm 0.06
\pm 0.01$ \\
$\Gamma_\perp/\Gamma = |A_\perp|^2$ & $0.23 \pm 0.19 \pm 0.04$ &
$0.13^{+0.12}_{-0.09} \pm 0.06$ \\ \hline
\end{tabular}
\end{center}
\end{table}

(2) A recent ALEPH analysis \cite{ALBs} of the decays $B_s \to D_s^{(*)+}
D_s^{(*)-}$ finds that the decay to pairs of vector mesons occurs in
mostly even partial waves, so that the lifetime in this mode probes
that of the CP-even mass eigenstate, which turns out to be $B_L \simeq (B_s +
\overline{B}_s)/\s$, giving $\tau_L = 1.27 \pm 0.33 \pm 0.07$ ps.  A similar
study of the flavor eigenstate finds $\tau(B_s) = 1.54 \pm 0.07$ ps.  Comparing
the two values, one finds $\Delta \Gamma/\Gamma = (25^{+21}_{-14})\%$.
This is just one facet of a combined analysis of results from
CDF, LEP, and SLD \cite{combBs} that concludes
$\Delta \Gamma/\Gamma = (16^{+8}_{-9})\%$, or $\Delta \Gamma/\Gamma < 31\%$
at 95\% c.l.

\section{$B$ decays to pairs of light mesons}

We have already noted in Table \ref{tab:PP} some branching ratios for $B$
decays to pairs of light pseudoscalar mesons.  Here we discuss these and
related processes involving one or two light vector mesons in more detail.

\subsection{Dominant processes in $B \to \pi \pi$ and $B \to K \pi$}

The decays $B \to \pi \pi$ and $B \to K \pi$ are rich in possibilities for
determining fundamental CKM parameters.  The process $B^0 \to \pi^+ \pi^-$
could yield the angle $\alpha$ in the absence of penguin amplitudes, whose
contribution must therefore be taken into account.  The process $B^0 \to K^+
\pi^-$ and related decays can provide information on the weak phase $\gamma$.

\begin{figure}
\centerline{\epsfysize=2.5in \epsffile{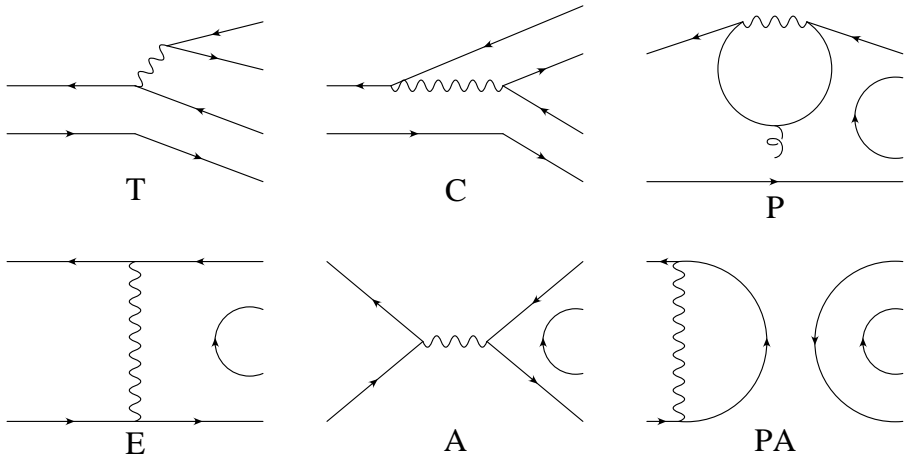}}
\caption{Graphs describing flavor-SU(3) invariant amplitudes for the decays of
$B$ mesons to pairs of light flavor-octet pseudoscalar mesons.  (a) ``Tree''
($T$); (b) ``Color-suppressed'' ($C$); (c) ``Penguin'' ($P$); (d) ``Exchange
($E$); (e) ``Annihilation'' ($A$); (f) ``Penguin annihilation'' ($PA$).}
\label{fig:gph}
\end{figure}

In order to discuss such decays in a unified way, we shall employ a
flavor-SU(3) description using a graphical representation.\cite{Chau,GHLR}
This language is equivalent to tensorial methods.\cite{Zepp,SW}  The
graphs are shown in Fig.~\ref{fig:gph}.  They constitute an over-complete
set; all processes of the form $B \to PP$, where $P$ is a light pseudoscalar
meson belonging to a flavor octet, are described by only 5 independent
linear combinations of these.

The graphical technique allows one to check a result
for $B \to \pi \pi$ which can be obtained using isospin invariance.
The subprocess $\bar b \to \bar u u \bar d$ can change isospin by 1/2 or 3/2
units.  The $J=0$ $\pi \pi$ final state, by virtue of Bose statistics, must
have even isospin:  $I=0,2$.  Thus there are only two invariant amplitudes
in the problem, one with $\Delta I = 1/2$ leading to $I_{\pi \pi} = 0$ and
one with $\Delta I = 3/2$ leading to $I_{\pi \pi} = 2$.  Hence the amplitudes
for the three decays $B^0 \to \pi^+ \pi^-$, $B^+ \to \pi^+ \pi^0$, and
$B^0 \to \pi^0 \pi^0$ obey one linear relation.  In the graphical
representation they are
\bea
A(B^0 \to \pi^+ \pi^-) & = & - (T+P)~~, \\
A(B^+ \to \pi^+ \pi^0) & = & -(T+C)/\s~~,\\
A(B^0 \to \pi^0 \pi^0) & = & (P-C)/\s~~,
\eea
leading to the relation $A(B^0 \to \pi^+ \pi^-) = \s A(B^+ \to \pi^+ \pi^0)
- \s A(B^0 \to \pi^0 \pi^0)$.  Measurement of the rates for these processes
and their charge-conjugates allows one to separate the penguin and tree
contributions from one another and to obtain information on the CKM phase
$\alpha$.  The only potential drawback of this method is that the branching
ratio for $A(B^0 \to \pi^0 \pi^0)$ is expected to be small: of order $10^{-6}$.

% This is Figure 7

One can use $B \to K \pi$ and flavor SU(3) to evaluate the penguin
contribution to $B \to \pi \pi$.\cite{SilWo} The decay $B \to \pi \pi$
appears to be dominated by the tree amplitude while $B \to K \pi$ appears to
be dominated by the penguin:
\beq
\left| \frac{({\rm Tree})_{K \pi}}{({\rm Tree})_{\pi \pi}} \right|
\simeq \left| \frac{f_K V_{us}}{f_\pi V_{ud}} \right| \simeq \left|
\frac{({\rm Pen})_{\pi \pi}}{({\rm Pen})_{K \pi}} \right| \simeq
\left| \frac{V_{td}}{V_{ts}} \right| \simeq \frac{1}{4}~~~.
\eeq
Many other applications of flavor SU(3) to $B \to K \pi$ decays have been
made subsequently.\cite{Kpi}  We shall discuss the results of one relatively
recent example.\cite{Bskpi}

\subsection{Measuring $\gamma$ with $B \to K \pi$ decays}

The Fermilab Tevatron and the CERN Large Hadron Collider (LHC) will produce
large numbers of $\pi^+ \pi^-$, $\pi^\pm K^\mp$, and $K^+ K^-$ pairs from
neutral non-strange and strange $B$ mesons.  Each set of decays has its
own distinguishing features.

The processes $B^0 \to K^+ K^-$ and $B_s \to \pi^+ \pi^-$ involve only
the spectator-quark amplitudes $E$ and $PA$, and thus should be suppressed.
They are related to one another by a flavor SU(3) ``U-spin'' reflection $s
\leftrightarrow d$ \cite{MGU} and thus the ratio of their rates should be the
ratio of the corresponding squares of CKM elements.

The decays $B^0 \to \pi^+ \pi^-$ and $B_s \to K^+ K^-$ also are related to
each other by a U-spin reflection.  Time-dependent studies of both processes
allow one to separate strong interaction and weak interaction information
from one another and to measure the angle $\gamma$.\cite{DuFl}  This appears
to be a promising method for Run II at the Fermilab Tevatron.\cite{Wurt}

The decays of non-strange and strange neutral $B$ mesons to $K^\pm \pi^\mp$
provide another source of information on $\gamma$ \cite{Bskpi} when combined
with information on $B^+ \to \ko \pi^+$. The rate for this process is predicted
to be the same as that for $B^- \to \ok \pi^-$, providing a consistency check.
We consider the amplitudes $T$ and $P$ with relative weak phase $\gamma$
and relative strong phase $\delta$, neglecting the amplitudes $E$, $A$, and
$PA$ which are expected to be suppressed relative to $T$ and $P$ by factors
of $f_B/m_B$.  Then we find (letting $T$ and $P$ stand for magnitudes)
\bea
A(\bo \to K^+ \pi^-) & = & - [P + T e^{i(\gamma + \delta)} ]~~, \\
A(B^+ \to \ko \pi^+) & = & P~~, \\
A(B_s \to \pi^+ K^-) & = & \tl P - \frac{1}{\tl} T e^{i(\gamma+\delta)}~~, \\
\eea
with amplitudes for the charge-conjugate processes given by $\gamma \to
- \gamma$.  Here $\tl \equiv |V_{us}/V_{ud}| = |V_{cd}/V_{cs}| = \tan
\theta_c \simeq 0.226$.  In the penguin amplitude the top quark has been
integrated out and unitarity used to replace $V^*_{tb} V_{tq}$ by $-V^*_{cb}
V_{cq} - V^*_{ub} V_{uq}$.  The term $-V^*_{cb}V_{cq}$ is the dominant
contribution to $P$, while $- V^*_{ub} V_{uq}$ is incorporated into $T$.

We define the charge-averaged ratios:
\beq
R \equiv \frac{\Gamma(B^0 \to K^+ \pi^-) + \Gamma(\bar B^0 \to K^- \pi^+)}
{\Gamma(B^+ \to K^0 \pi^+) + \Gamma(B^- \to \bar K^0 \pi^-)}~~~,
\eeq
\beq
R_s \equiv \frac{\Gamma(B_s \to K^- \pi^+) + \Gamma(\bar B_s \to K^+ \pi^-)}
{\Gamma(B^+ \to K^0 \pi^+) + \Gamma(B^- \to \bar K^0 \pi^-)}~~~,
\eeq
and CP-violating rate (pseudo-)asymmetries:
\beq
A_0 \equiv \frac{\Gamma(B^0 \to K^+ \pi^-) -\Gamma(\bar B^0 \to K^- \pi^+)}
{\Gamma(B^+ \to K^0 \pi^+) + \Gamma(B^- \to \bar K^0 \pi^-)}~~~,
\eeq
\beq
A_s \equiv \frac{\Gamma(B_s \to K^- \pi^+) - \Gamma(\bar B_s \to K^+ \pi^-)}
{\Gamma(B^+ \to K^0 \pi^+) + \Gamma(B^- \to \bar K^0 \pi^-)}~~~.
\eeq
and let $r \equiv T/P$.  We find
\beq \label{eqn:R}
R = 1 + r^2 + 2 r \cos \delta \cos \gamma~~~,
\eeq
\beq \label{eqn:Rs}
R_s = \tl^2 + (r/\tl)^2 - 2 r \cos \delta \cos \gamma~~~,
\eeq
\beq \label{eqn:asym}
A_0 = - A_s = -2r \sin \gamma \sin \delta~~~.
\eeq
The relation $A_0 = -A_s$ may be used to test the assumption of flavor
SU(3) symmetry, while the remaining three equations may be solved for the
three unknowns $r$, $\gamma$, and $\delta$.  An error of $10^\circ$
on $\gamma$ seems feasible.  (A small correction associated with the
above approximation to the penguin graph also may be applied.\cite{CW})

\subsection{Decays with $\eta$ and $\eta'$ in the final state}

\begin{figure}
\centerline{\epsfysize=1.5in \epsffile{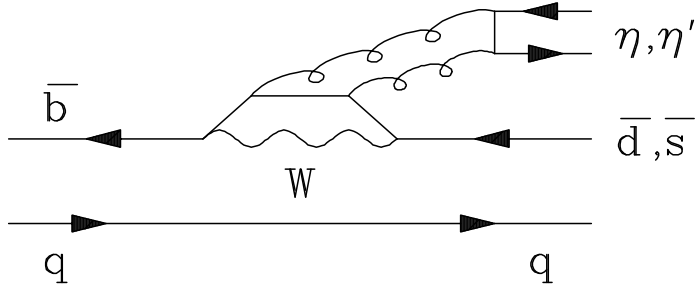}}
\caption{Singlet penguin diagram, important for $B \to PP$ processes in
which one of the pseudoscalar mesons $P$ is $\eta$ or $\eta'$.}
\label{fig:sp}
\end{figure}

The physical $\eta$ and $\eta'$ are mixtures of the flavor octet state
$\eta_8 \equiv (2 s \bar s - u \bar u - d \bar d)/\sx$ and the flavor
singlet $\eta_1 \equiv (u \bar u + d \bar d + s \bar s)/\st$.  This mixing
is tested in many decays, such as $(\eta,\eta') \to \gamma \gamma$, $(\rho,
\omega, \phi) \to \eta \gamma$, $\eta' \to (\rho,\omega) \gamma$, etc.  The
result \cite{etamix} is that the $\eta$ is mostly an octet and the $\eta'$
mostly a singlet, with one frequently-employed approximation \cite{Chau,etapx}
corresponding to an octet-singlet mixing angle of $19^\circ$:
\beq
\eta \simeq \frac{1}{\st}(s \bar s - u \bar u - d \bar d)~~,~~~
\eta' \simeq \frac{1}{\sx}(u \bar u + d \bar d + 2 s \bar s)~~~,
\eeq
(A single mixing angle may not adequately describe the $\eta$--$\eta'$
system.\cite{Feld})
For a meson with a flavor singlet component, an amplitude in addition
to those depicted in Fig.\ \ref{fig:gph} corresponds to the ``singlet''
penguin diagram \cite{etapx} shown in Fig.\ \ref{fig:sp}.

% This is Figure 8

The CLEO Collaboration's large branching ratios for
$B \to \eta' K$: \cite{CLetap}
\beq
{\cal B}(B^+ \to \eta' K^+) = (80^{+10}_{-9} \pm 7) \times 10^{-6}~~,
~~~{\cal B}(\bo \to \eta' \ko) = (89^{+18}_{-16} \pm 9) \times 10^{-6}~~~,
\eeq
with only upper limits for $K \eta$ production,
indicate the presence of a substantial ``singlet'' penguin contribution,
and constructive interference between nonstrange and strange quark
contributions of $\eta'$ to the {\it ordinary} penguin amplitude $P$, as
suggested by Lipkin.\cite{HJLeta}
The corresponding decays to nonstrange final states, $B^+ \to \pi^+ \eta$ and
$B^+ \to \pi^+ \eta'$, are expected to have large CP-violating asymmetries,
since several weak amplitudes in these processes are of comparable
magnitude.\cite{etapx} Moreover, CLEO sees
\bea
{\cal B}(B^+ \to \eta K^{*+}) & = & (26.4^{+9.6}_{-8.2} \pm 3.3)
\times 10^{-6}~~~, \\
{\cal B}(\bo \to \eta K^{*0}) & = & (13.8^{+5.5}_{-4.6} \pm 1.6)
\times 10^{-6}~~~,
\eea
with only upper limits for $K^* \eta'$ production.  These results favor the
{\it opposite} signs for nonstrange and strange components of the $\eta$,
again in accord with predictions.\cite{HJLeta}

Much theoretical effort has been expended on attempts to understand the
magnitude of the ``singlet'' penguin diagrams,\cite{spth} but they appear
to be more important than one would estimate using pertubative QCD.
 
\subsection{One vector meson and one pseudoscalar}

The decays $B \to VP$, where $V$ is a vector meson and $P$ a pseudoscalar,
are characterized by twice as many invariant amplitudes of flavor SU(3)
as the decays $B \to PP$, since either the vector meson or the pseudoscalar
can contain the spectator quark.  We can label the corresponding amplitudes
by a subscript $V$ or $P$ to denote the type of meson containing the
spectator.  A recent analysis within the graphical framework uses data to
specify amplitudes.\cite{GRVP}  Alternatively, one can incorporate models for
form factors into calculations based on factorization.\cite{Hou,He}  An
interesting possibility suggested in both these approaches is that the large
branching ratio ${\cal B}(\bo \to K^{*+} \pi^-)$ may suggest constructive
tree-penguin interference, implying $\gamma \ge 90^\circ$.

The tree amplitude in $\bo \to K^{*+} \pi^-$ is proportional to $V^*_{ub}
V_{us}$, with weak phase $\gamma$, while the penguin amplitude is proportional
to $V^*_{tb} V_{ts}$, with weak phase $\pi$.  The relative weak phase between
these two amplitudes is then $\gamma - \pi$, which leads to constructive
interference if the strong phase difference between the tree and penguin
amplitudes is small and if $\Gamma > \pi/2$.  This could help explain why
${\cal B}(\bo \to K^{*+} \pi^-)$ seems to exceed $2 \times 10^{-5}$
while the pure penguin process $B \to \phi K$ corresponds to a branching
ratio of only $(6.2^{+2.0+0.7}_{-1.8-1.7}) \times 10^{-6}$.\cite{CL2K,CLVP}

A similar tree-penguin interference can occur in $B^0 \to \pi^+ \pi^-$.  The
tree amplitude is proportional to $V^*_{ub} V_{ud}$, with weak phase $\gamma$,
while the penguin is proportional to $V^*_{tb} V_{td}$, with weak phase
$- \beta$.  The relative weak phase is then $\gamma + \beta = \pi - \alpha$.
One expects destructive interference if the final strong phase difference is
small and $\alpha < \pi/2$.  This could help explain why $B(\bo \to \pi^+
%U           | |     | |
\pi^-)$ is $(4.4 \pm 0.9) \times 10^{-6}$ while the tree contribution alone,
estimated (for example) from $B \to \pi \ell \nu$,\cite{GR98} would lead
to a branching ratio around $10^{-5}$.

A global fit of $B \to PP$ and $B\to VP$ data \cite{Hou} based on factorization
and models for form factors leads to $\gamma = (114^{+25}_{-23})^{\circ}$,
which just barely clips the corner of the allowed $(\rho,\eta)$ region.  If
valid, this result implies that we should see $\Delta m_s$ near its present
lower bound.

\subsection{Two vector mesons}

No modes with pairs of light vector mesons have been identified conclusively
yet. The existence of three partial waves (S, P, D) for such processes as $\bo
\to \phi K^{*0}$ means that helicity analyses can in principle detect the
presence of final-state interactions (as in the case of $B \to J/\psi K^*$).
It is not clear, however, whether such final-state phases are
relevant to the case of greatest interest, in which two different channels
are ``fed'' by different {\it weak} processes such as $T$ and $P$ amplitudes.
Some further information obtainable from angular distributions in $B
\to VV$ decays has been noted.\cite{Chiang}

\subsection{Testing flavor SU(3)}

The asymmetry prediction $A_s = - A_0$ for $B_s \to \bar K \pi$ vs.
$B \to K \pi$, mentioned above, is just one of a number of U-spin relations
\cite{MGU} testable via
$B_s$ decays, which will first be studied in detail at hadron colliders.
One expects the assumption of flavor SU(3), and in particular the equality of
final-state phases for non-strange and strange $B$ final states, to be more
valid for $B$ decays than for charm decays, where resonances still are
expected to play a major role.
 
\section{The role of penguins}

\subsection{Estimates of magnitudes}

Perturbative calculations of penguin contributions to processes such as
$B \to K \pi$, where they seem to be dominant, fall short of actual
measurements.\cite{Ciu}  Phenomenological fits indicate no
suppression by a factor of $\alpha_s/4 \pi$ despite the presence of a loop and
a gluon.  One possible explanation is the
presence of a $c \bar c$ loop with substantial enhancement from on-shell
states, equivalent to strong rescattering from such states as
$D_s \bar D$ to charmless meson pairs.  If this is indeed the case,
penguin amplitudes could have different final-state phases from tree
amplitudes, enhancing the possibility of observing direct CP violation.

\subsection{Electroweak penguins}

% This is Figure 9
\begin{figure}
\centerline{\epsfysize=1.5in \epsffile{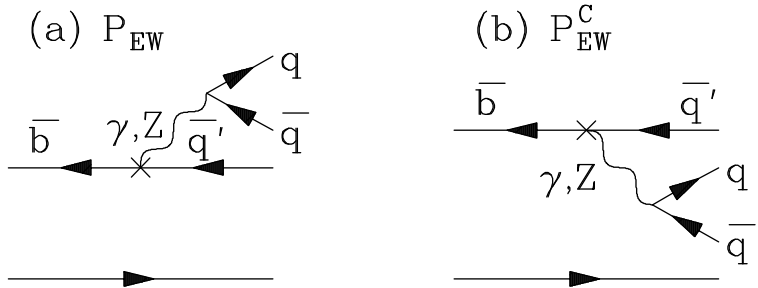}}
\caption{Electroweak penguin (EWP) diagrams.  (a) Color-favored ($P_{EW}$);
(b) Color-suppressed ($P^c_{EW}$).}
\label{fig:ewp}
\end{figure}

When the gluon in a penguin diagram is replaced by a (real or virtual) photon
or a virtual $Z$ which couples to a final $q \bar q$ pair,
the process $\bar b \to (\bar d~{\rm or}~\bar s) q \bar q$
is no longer independent of the flavor ($u,d,s$) of $q$.
Instead, one has contributions in which the $u \bar u$ pair is treated
differently from the $d \bar d$ or $s \bar s$ pair.  A {\it color-favored}
electroweak penguin amplitude $P_{EW}$ [Fig.\ \ref{fig:ewp}(a)]
involves the pair appearing in the same neutral meson (e.g., $\pi^0$), while a
{\it color-suppressed} amplitude $P^c_{EW}$ [Fig.\ \ref{fig:ewp}(b)]
involves each member of the pair appearing in a different meson.
 
One may parametrize electroweak penguin (EWP) amplitudes
by contributions proportional to the quark charge, sweeping other terms into
the gluonic penguin contributions.  One then finds that the EWP
terms in a flavor-SU(3) description may be combined as follows with
the terms $T$, $C$, $P$, and $S$ (the ``singlet'') penguin: \cite{GHLRP}
\bea
T \to t & \equiv & T + P^c_{EW}~~,~~P \to p \equiv P - \frac{1}{3}P^c_{EW}~~,\\
C \to c & \equiv & C + P_{EW}~~,~~S \to s \equiv S - \frac{1}{3}P_{EW}~~.
\eea
The flavor-SU(3) description holds as before, but weak phases now can differ
from their previous values as a result of the EWP contributions.

\begin{figure}
\centerline{\epsfysize=2in \epsffile{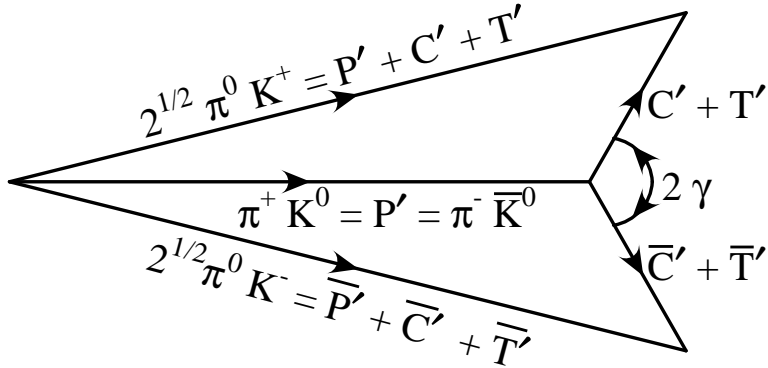}}
\caption{Amplitude triangles for determining the weak phase $\gamma$.  These
are affected by electroweak penguin contributions, as described in the text.}
\label{fig:gam}
\end{figure}

One early application of flavor SU(3) which turns out to be significantly
affected by EWP contributions is the attempt to learn the weak phase $\gamma$
from information on the decays $B^+ \to K^+ \pi^0$, $B^+ \to \ko \pi^+$,
$B^+ \to \pi^+ \pi^0$, and the corresponding charge-conjugate
decays.\cite{GLRPRL}  The amplitude construction is illustrated in Fig.\
\ref{fig:gam}.  The primes on the amplitudes refer to the fact that they
describe strangeness-changing ($|\Delta S| = 1$) transitions. The corresponding 
$\Delta S = 0$ amplitudes are unprimed.

% This is Figure 10

The amplitudes in Fig.\ \ref{fig:gam} form two triangles, whose sides
labeled $C' + T'$ and $\bar C' + \bar T'$ form an angle $2 \gamma$ with
respect to one another.  (There will be a discrete ambiguity corresponding
to flipping one of the triangles about its base.)  One estimates the
lengths of these two sides using flavor SU(3) from the amplitudes
$A(B^+ \to \pi^+ \pi^0) = (C + T)/\s$ and $A(B^- \to \pi^- \pi^0) = (\bar C
+ \bar T)/\s$.

In the presence of electroweak penguin contributions this simple analysis
must be modified, since there are important additional contributions when
we replace $C'+T' \to c'+t'$ and $\bar C' + \bar T' \to \bar c' + \bar
t'$.\cite{DH}  The culprit is the $C'$ amplitude, which is associated with a
color-favored electroweak penguin.  It was noted subsequently \cite{NR}
that since the $C'+T'$ amplitude corresponds to isospin $I(K \pi) = 3/2$
for the final state, the strong-interaction phase of its EWP contribution is
the same as that of the rest of the $C'+T'$ amplitude, permitting the
calculation of the EWP correction.  The result is that
\beq
A[I(K \pi) = 3/2] \sim {\rm const} \times (e^{i \gamma} - \delta_{EW})~~~,
\eeq
where the phase in the first term is Arg($V^*_{ub}V_{us})$ and the second term
is estimated to be $\delta_{EW} = 0.64 \pm 0.15$ when SU(3)-breaking effects
are included.

Any deviation of the ratio $2 \Gamma(B^+ \to K^+ \pi^0)/\Gamma(B^+ \to \ko
\pi^+)$ from 1 can signify interference of the $C'+T'$ amplitude with the
dominant $P'$ amplitude and hence can provide information on $\gamma$.  The
present value for this ratio, based on the branching ratios in Table 
\ref{tab:PP}, is $1.27 \pm 0.47$, compatible with 1.

The triangle construction (with its generalization to include
EWP contributions) avoids the need to evaluate strong phases.  Other studies of
calculable electroweak penguin effects have been made.\cite{GP}  Good use of
these results will require 100 fb$^{-1}$ at an $e^+ e^-$ collider or
$10^8$ produced $B \bar B$ pairs.

\section{Final-state interactions}

It is crucial to understand final-state strong phases in
order to anticipate direct CP-violating rate asymmetries and to check whether
assumptions about the smallness of amplitudes involving the spectator quark
are correct.  The decay $B^+ \to \ko \pi^+$
in the na\"{\i}ve diagrammatic approach is expected to be dominated by the
penguin diagram with no tree contribution.  The penguin weak phase would be
Arg($V^*_{tb}V_{ts}) = \pi$. The phase of the annihilation amplitude $A$, which
is expected to be suppressed by a factor of $\lambda^2 f_B/m_B$ and hence
should be unimportant, should be Arg($V^*_{ub}V_{us}) = \gamma$.  This
implies a very small CP-violating rate asymmetry between $B^+ \to \ko \pi^+$
and $B^- \to \ok \pi^-$, much smaller than in cases where $T$ and $P$
amplitudes can interfere such as $\bo \to K^+ \pi^-$.

\begin{table}[h]
\caption{CP-violating rate asymmetries for several $B \to K \pi$
processes. \label{tab:asy}}
\begin{center}
\begin{tabular}{c c c} \hline
Mode & Signal events & ${\cal A}_{CP}$ \\ \hline
$K^+ \pi^-$ & $80^{+12}_{-11}$ & $-0.04 \pm 0.16$ \\
$K^+ \pi^0$ & $42.1^{+10.9}_{-9.9}$ & $-0.29 \pm 0.23$ \\
$K_S \pi^+$ & $25.2^{+6.4}_{-5.6}$ & $0.18 \pm 0.24$ \\
$K^+ \eta'$ & $100^{+13}_{-12}$ & $0.03 \pm 0.12$ \\
$\omega \pi^+$ & $28.5^{+8.2}_{-7.3}$ & $-0.34 \pm 0.25$ \\ \hline
\end{tabular}
\end{center}
\end{table}

The current data do not exhibit significant CP asymmetries in any
modes.\cite{Chen}  In Table
\ref{tab:asy} we summarize some recently report CP asymmetries, defined as
\beq
{\cal A}_{CP} \equiv \frac{\Gamma(\bar B \to \bar f) - \Gamma(B \to f)}
{\Gamma(\bar B \to \bar f) + \Gamma(B \to f)}~~~.
\eeq
The asymmetry in the mode $K_S \pi^\pm$ is no more or less significant than
in other modes where ${\cal A}_{CP} \ne 0$ could be expected.  How could we
tell whether the amplitude $A$ is suppressed by as much as we expect in the
na\"{\i}ve approach?

\subsection{Rescattering}

Rescattering from tree processes (such as those in Fig.\ \ref{fig:kpi1} or
Fig.\ \ref{fig:kpi2} contributing to $B^+ \to K^+ \pi^0$) could amplify
the effective $A$ amplitude in $B^+ \to K^0 \pi^+$, removing the suppression
factor of $f_B/m_B$.
The tree amplitude for
$B^+ \to K^+ \pi^0$ should be proportional to $V^*_{ub} V_{us}$ (as in the
$A$ amplitude), but the magnitude of the rescattering amplitude for $K^+ \pi^0
\to \ko \pi^+$ (in an S-wave) is unknown at the center-of-mass energy of
$m_B c^2 = 5.28$ GeV.

\subsection{A useful SU(3) relation}

A sensitive test for the presence of an enhanced $A$ amplitude has been
proposed, \cite{FalkU} utilizing the U-spin symmetry $d \leftrightarrow s$
of flavor SU(3).  Under this transformation, the $\bar b \to \bar s$ penguin
diagram contributing to $B^+ \to \ko \pi^+$ is transformed into the $\bar b \to
\bar d$ penguin contributing to $B^+ \to \ok K^+$, suppressed by a relative
factor of $|V^*_{tb} V_{td}/V^*_{tb} V_{ts}| \simeq \lambda$, while the
annihilation diagram contributing to $B^+ \to \ko \pi^+$ is transformed into
that contributing to $B^+ \to \ok K^+$, {\it enhanced} by a relative factor of
$|V^*_{ub} V_{ud}/V^*_{ub} V_{us}| \simeq 1/\lambda$.  Thus the relative
effects of the ``annihilation'' amplitude should be stronger by a factor of
$1/\lambda^2$ in $B^+ \to \ok K^+$ than in $B^+ \to \ko \pi^+$.  Even if these
effects are not large enough to significantly influence the decay rate, they
could well influence the predicted decay asymmetry. 

\subsection{The process $\bo \to K^+ K^-$}

A process which should be {\it dominated} by interactions involving the
spectator quark is $\bo \to K^+ K^-$.\cite{GRres}
Only the exchange ($E$) and penguin annihilation ($PA$) graphs in Fig.\
\ref{fig:gph} contribute to this decay.

The exchange ($E$) amplitude should be proportional to $(f_B/m_B)V^*_{ub}
V_{ud}$, and the penguin annihilation amplitude should be suppressed by
further powers of $\alpha_s$, in a na\"{\i}ve approach.  The expected
branching ratio if the $E$ amplitude dominates should be less than $10^{-7}$.
However, if rescattering is important, the $K^+ K^-$ final state could be
``fed'' by the process $\bo \to \pi^+ \pi^-$, whose amplitude is proportional
to $T+P$.  Present experimental limits place only the 90\% c.l. upper bound
${\cal B}(\bo \to K^+ K^-) < 1.9 \times 10^{-6}$.\cite{CLEOkpi} 

\subsection{Critical remarks}

Some estimates of rescattering are based on Regge pole methods.\cite{RP}
These may not apply to low partial waves at energies of $m_b c^2 = 5.28$
GeV.  Regge poles have proven phenomenologically successful primarily
for ``peripheral'' partial waves $\ell \sim k_{\rm c.m.} \times (R \sim 1
{\rm~fm})$.\cite{HHPW}
 
\section{Topics not covered in detail}

\subsection{Measurement of $\gamma$ using $B^\pm \to D K$ decays}

The self-tagging decays $B^\pm \to D^0 K^\pm$, $B\pm \to \od K^\pm$, and
$B^\pm \to D_{CP} K^\pm$, where $D_{CP}$ is a CP eigenstate, permit one
to perform a triangle construction very similar to that in Fig.\ \ref{fig:gam}
to extract the weak phase $\gamma$.\cite{GW}  However, the interference
of the Cabibbo-favored decay $D^0 \to K^- \pi^+$ and the
doubly-Cabibbo-suppressed decay $D^0 \to K^+ \pi^-$ introduces an important
subtlety in this method, which has been addressed.\cite{but}

\subsection{Dalitz plot analyses}

The likely scarcity of the decay $B^0 \to \pi^0 \pi^0$ (see Sec.\ 10) may
be an important limitation in the method proposed \cite{GrL} to extract
the weak phase $\alpha$ from $B \to \pi \pi$ decays using an isospin analysis.
It has been suggested \cite{SQ} that one study instead the isospin structure of
the decays $B \to \rho \pi$, since at least some of these processes occur with
greater branching ratios than the corresponding $B \to \pi \pi$ decays.
One must thus measure time-dependences and total rates for the processes
$(\bo~{\rm or}~\ob) \to (\rho^\pm \pi^\mp, \rho \pi^0)$.  A good deal of useful
information, in fact, can be learned just from the time-integrated
rates.\cite{QS}

\subsection{CP violation in $\bo$--$\ob$ mixing}

The standard model of CP violation predicts that the number of same-sign
dilepton pairs due to $\bo$--$\ob$ mixing should be nearly the same for
$\ell^+ \ell^+$ and $\ell^- \ell^-$.  By studying such pairs it is possible
to test not only this prediction, but also the validity of CPT invariance.
The OPAL Collaboration \cite{OPCPT} parametrizes neutral non-strange $B$ mass
eigenstates as
\bea
\ket{B_1} & = & \frac{(1 + \epsilon_B + \delta_B) \ket{\bo}
                    + (1 - \epsilon_B - \delta_B) \ket{\ob}}
                {\sqrt{2(1 + |\epsilon_B + \delta_B|^2)}}~~,\\
\ket{B_2} & = & \frac{(1 + \epsilon_B - \delta_B) \ket{\bo}
                    - (1 - \epsilon_B + \delta_B) \ket{\ob}}
                {\sqrt{2(1 + |\epsilon_B -  \delta_B|^2)}}~~~
\eea
and finds (allowing for CPT
violation) Im($\delta_B) = -0.020 \pm 0.016 \pm 0.006$, Re($\epsilon_B) =
-0.006 \pm 0.010 \pm 0.006$.  Enforcing CPT invariance, they find
Re($\epsilon_B) = 0.002 \pm 0.007 \pm 0.003$.  A recent CLEO study
\cite{CLEOeb} finds Re($\epsilon_B)/(1 + |\epsilon_B|^2) = 0.0035 \pm
0.0103 \pm 0.0015$ under similar assumptions.  The standard model predicts
$|p/q| \simeq 1$ and hence Re($\epsilon_B) \simeq 0$.
   
\section{What if the CKM picture doesn't work?}

\subsection{Likely accuracy of future measurements}

\begin{figure}
\centerline{\epsfysize=2.5in \epsffile{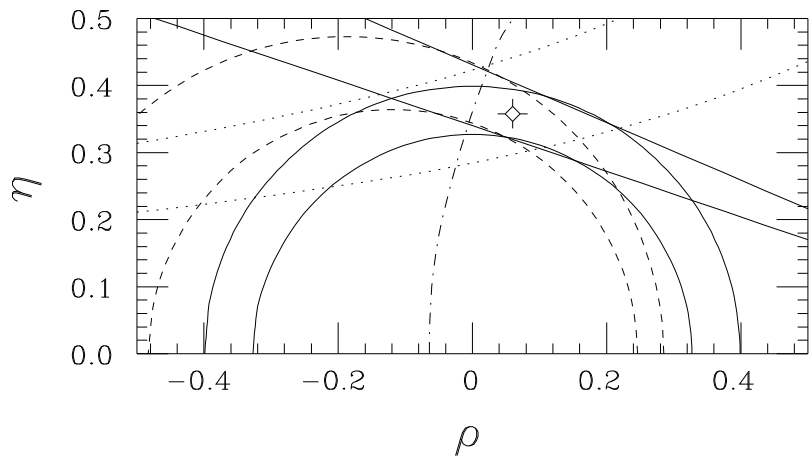}}
\caption{Plot in the $(\rho,\eta)$ plane of anticipated $\pm 1 \sigma$
constraints on CKM
parameters in the year 2003.  Solid curves: $|V_{ub}/V_{cb}|$; dashed lines:
constraint on $|V_{ub}/V_{td}|$ by combining measurement of ${\cal B}(B^+ \to
\tau^+ \nu_\tau)$ with $\bo$--$\ob$ mixing; dotted lines: constraint due to
$\epsilon_K$ (CP-violating $\ko$--$\ok$ mixing); dash-dotted line: limit
due to $x_s$; solid rays:  measurement of $\sin 2 \beta$ to $\pm 0.06$.}
\label{fig:fut}
\end{figure}

It's useful to anticipate how our knowledge of the Cabibbo-Kobayashi-Maskawa
matrix might evolve over the next few years.\cite{JRlat,fut}  With $\sin(2
\beta)$
measured in $\bo \to J/\psi K_S$ decays to an accuracy of $\pm 0.06$ (the
BaBar goal with 30 fb$^{-1}$ \cite{BaBarbk}), errors on $|V_{ub}/V_{cb}|$
reduced to 10\%, strange-$B$ mixing bounded by $x_s = \Delta m_s/\Gamma_s
> 20$ (the present bound is already better than this!), and ${\cal B}(B^+ \to
\tau^+ \nu_\tau)$ measured to $\pm 20\%$ (giving $f_B|V_{ub}$, or $|V_{ub}/\
V_{td}|$ when combined with $\bo$--$\ob$ mixing), one finds the result
shown in Fig.\ \ref{fig:fut}.

The anticipated $(\rho,\eta)$ region is quite restricted, leading to the
likelihood that if physics beyond the standard model is present, it will
show up in such a plot as a contradiction among various measurements.  What
could be some sources of new physics?

% This is Figure 11

\subsection{Possible extensions}

Some sources of effects beyond the standard model which could show up first
in studies of $B$ mesons include:

\begin{itemize}

\item Supersymmetry (nearly everyone's guess) (see Murayama's
lectures \cite{Mura});

\item Flavor-changing effects from extended models of dynamical electroweak
symmetry breaking (mentioned, e.g., by Chivukula \cite{Chiv});

\item Mixing of ordinary quarks with exotic ones, as in certain versions of
grand unified theories.\cite{NS}

\end{itemize}

Typical effects show up most prominently in mixing (particularly $\ko$--$\ok$
and $\bo$--$\ob$),\cite{mix} but also could appear in penguin processes
such as $B \to \phi K$.\cite{GrWo}

\section{Summary}

We are entering an exciting era of precision $B$ physics.  With experimental
and theoretical advances occurring on many fronts, we have good reason to
hope for surprises in the next few years.  If, however, the present picture
survives such stringent tests, we should turn our attention to the more
fundamental question of where the CKM matrix (as well as the quark masses
themselves!) actually originates.

\section*{Acknowledgments}

I would like to thank Prof.\ K. T. Mahanthappa for his directorship of the
TASI-2000 Summer School and for his gracious hospitality in Boulder.  These 
lectures were prepared in part during visits to DESY and Cornell; I thank
colleagues there for hospitality and for the chance to discuss some of the
subject material with them.  These lectures grew out of long-term
collaborations with Amol Dighe, Isard Dunietz, and Michael Gronau.  I am
grateful to them and to others, including O. F. Hern\'andez, H. J. Lipkin,
D. London, and M. Neubert, for many pleasant interactions on these subjects.
This work was supported in part by the United States Department of Energy
through Grant No.\ DE FG02 90ER40560.

\section*{References}

% Journal and other miscellaneous abbreviations for references
\def \ajp#1#2#3{Am.\ J. Phys.\ {\bf#1}, #2 (#3)}
\def \apny#1#2#3{Ann.\ Phys.\ (N.Y.) {\bf#1}, #2 (#3)}
\def \app#1#2#3{Acta Phys.\ Polonica {\bf#1}, #2 (#3)}
\def \arnps#1#2#3{Ann.\ Rev.\ Nucl.\ Part.\ Sci.\ {\bf#1}, #2 (#3)}
\def \art{and references therein}
\def \cmts#1#2#3{Comments on Nucl.\ Part.\ Phys.\ {\bf#1}, #2 (#3)}
\def \cn{Collaboration}
\def \cp89{{\it CP Violation,} edited by C. Jarlskog (World Scientific,
Singapore, 1989)}
\def \econf#1#2#3{Electronic Conference Proceedings {\bf#1}, #2 (#3)}
\def \efi{Enrico Fermi Institute Report No.\ }
\def \epjc#1#2#3{Eur.\ Phys.\ J.\ C {\bf#1}, #2 (#3)}
\def \f79{{\it Proceedings of the 1979 International Symposium on Lepton and
Photon Interactions at High Energies,} Fermilab, August 23-29, 1979, ed. by
T. B. W. Kirk and H. D. I. Abarbanel (Fermi National Accelerator Laboratory,
Batavia, IL, 1979}
\def \hb87{{\it Proceeding of the 1987 International Symposium on Lepton and
Photon Interactions at High Energies,} Hamburg, 1987, ed. by W. Bartel
and R. R\"uckl (Nucl.\ Phys.\ B, Proc.\ Suppl., vol. 3) (North-Holland,
Amsterdam, 1988)}
\def \ib{{\it ibid.}~}
\def \ibj#1#2#3{~{\bf#1}, #2 (#3)}
\def \ichep72{{\it Proceedings of the XVI International Conference on High
Energy Physics}, Chicago and Batavia, Illinois, Sept. 6 -- 13, 1972,
edited by J. D. Jackson, A. Roberts, and R. Donaldson (Fermilab, Batavia,
IL, 1972)}
\def \ijmpa#1#2#3{Int.\ J.\ Mod.\ Phys.\ A {\bf#1}, #2 (#3)}
\def \ite{{\it et al.}}
\def \jhep#1#2#3{JHEP {\bf#1}, #2 (#3)}
\def \jpb#1#2#3{J.\ Phys.\ B {\bf#1}, #2 (#3)}
\def \lg{{\it Proceedings of the XIXth International Symposium on
Lepton and Photon Interactions,} Stanford, California, August 9--14, 1999,
edited by J. Jaros and M. Peskin (World Scientific, Singapore, 2000)}
\def \lkl87{{\it Selected Topics in Electroweak Interactions} (Proceedings of
the Second Lake Louise Institute on New Frontiers in Particle Physics, 15 --
21 February, 1987), edited by J. M. Cameron \ite~(World Scientific, Singapore,
1987)}
\def \kaon{{\it Kaon Physics}, edited by J. L. Rosner and B. Winstein,
%U                              |
University of Chicago Press, 2001}
\def \kdvs#1#2#3{{Kong.\ Danske Vid.\ Selsk., Matt-fys.\ Medd.} {\bf #1}, No.\
#2 (#3)}
\def \ky{{\it Proceedings of the International Symposium on Lepton and
Photon Interactions at High Energy,} Kyoto, Aug.~19-24, 1985, edited by M.
Konuma and K. Takahashi (Kyoto Univ., Kyoto, 1985)}
\def \mpla#1#2#3{Mod.\ Phys.\ Lett.\ A {\bf#1}, #2 (#3)}
\def \nat#1#2#3{Nature {\bf#1}, #2 (#3)}
\def \nc#1#2#3{Nuovo Cim.\ {\bf#1}, #2 (#3)}
\def \nima#1#2#3{Nucl.\ Instr.\ Meth.\ A {\bf#1}, #2 (#3)}
\def \np#1#2#3{Nucl.\ Phys.\ {\bf#1}, #2 (#3)}
\def \npps#1#2#3{Nucl.\ Phys.\ Proc.\ Suppl.\ {\bf#1}, #2 (#3)}
\def \os{XXX International Conference on High Energy Physics, Osaka, Japan,
July 27 -- August 2, 2000}
\def \PDG{Particle Data Group, D. E. Groom \ite, \epjc{15}{1}{2000}}
\def \pisma#1#2#3#4{Pis'ma Zh.\ Eksp.\ Teor.\ Fiz.\ {\bf#1}, #2 (#3) [JETP
Lett.\ {\bf#1}, #4 (#3)]}
\def \pl#1#2#3{Phys.\ Lett.\ {\bf#1}, #2 (#3)}
\def \pla#1#2#3{Phys.\ Lett.\ A {\bf#1}, #2 (#3)}
\def \plb#1#2#3{Phys.\ Lett.\ B {\bf#1}, #2 (#3)}
\def \pr#1#2#3{Phys.\ Rev.\ {\bf#1}, #2 (#3)}
\def \prc#1#2#3{Phys.\ Rev.\ C {\bf#1}, #2 (#3)}
\def \prd#1#2#3{Phys.\ Rev.\ D {\bf#1}, #2 (#3)}
\def \prl#1#2#3{Phys.\ Rev.\ Lett.\ {\bf#1}, #2 (#3)}
\def \prp#1#2#3{Phys.\ Rep.\ {\bf#1}, #2 (#3)}
\def \ptp#1#2#3{Prog.\ Theor.\ Phys.\ {\bf#1}, #2 (#3)}
\def \rmp#1#2#3{Rev.\ Mod.\ Phys.\ {\bf#1}, #2 (#3)}
\def \rp#1{~~~~~\ldots\ldots{\rm rp~}{#1}~~~~~}
\def \si90{25th International Conference on High Energy Physics, Singapore,
Aug. 2-8, 1990}
\def \slc87{{\it Proceedings of the Salt Lake City Meeting} (Division of
Particles and Fields, American Physical Society, Salt Lake City, Utah, 1987),
ed. by C. DeTar and J. S. Ball (World Scientific, Singapore, 1987)}
\def \slac89{{\it Proceedings of the XIVth International Symposium on
Lepton and Photon Interactions,} Stanford, California, 1989, edited by M.
Riordan (World Scientific, Singapore, 1990)}
\def \smass82{{\it Proceedings of the 1982 DPF Summer Study on Elementary
Particle Physics and Future Facilities}, Snowmass, Colorado, edited by R.
Donaldson, R. Gustafson, and F. Paige (World Scientific, Singapore, 1982)}
\def \smass90{{\it Research Directions for the Decade} (Proceedings of the
1990 Summer Study on High Energy Physics, June 25--July 13, Snowmass, Colorado),
edited by E. L. Berger (World Scientific, Singapore, 1992)}
\def \tasi{{\it Testing the Standard Model} (Proceedings of the 1990
Theoretical Advanced Study Institute in Elementary Particle Physics, Boulder,
Colorado, 3--27 June, 1990), edited by M. Cveti\v{c} and P. Langacker
(World Scientific, Singapore, 1991)}
\def \yaf#1#2#3#4{Yad.\ Fiz.\ {\bf#1}, #2 (#3) [Sov.\ J.\ Nucl.\ Phys.\
{\bf #1}, #4 (#3)]}
\def \zhetf#1#2#3#4#5#6{Zh.\ Eksp.\ Teor.\ Fiz.\ {\bf #1}, #2 (#3) [Sov.\
Phys.\ - JETP {\bf #4}, #5 (#6)]}
\def \zpc#1#2#3{Zeit.\ Phys.\ C {\bf#1}, #2 (#3)}
\def \zpd#1#2#3{Zeit.\ Phys.\ D {\bf#1}, #2 (#3)}

\end{document}